\pdfoutput=1
\documentclass[galaxies,article,accept,moreauthors,pdftex,10pt,a4paper,usenatbib]{mdpi}

\firstpage{1}
\articlenumber{x}
\doinum{10.3390/------}
\pubvolume{xx}
\pubyear{2017}
\copyrightyear{2017}
\externaleditor{Academic Editors: Duncan A. Forbes and Ericson D. Lopez}
\history{Received: 1 August 2017; Accepted: 28 August 2017; Published: date}

\usepackage{graphicx}
\usepackage{sidecap}
\sidecaptionvpos{figure}{c}

\newcommand{\eprint}[2][]{{\tt\if!#1!#2\else#1:#2\fi}}

\usepackage{booktabs}
\usepackage{multirow}
\usepackage{soul}
\usepackage{microtype}

\Title{The Outer Halos of Very Massive Galaxies:
BCGs and their DSC in the Magneticum Simulations}

\Author{{Rhea-Silvia Remus}\,$^{1,}$*,
Klaus Dolag\,$^{1,2}$
and Tadziu L. Hoffmann\,$^1$}

\AuthorNames{Rhea-Silvia Remus, Klaus Dolag, \& Tadziu L. Hoffmann}

\address{%
$^{1}$ \quad
{Universit\"ats-Sternwarte M\"unchen},
Faculty of Physics,
Ludwig-Maximilians-Universit\"at,
Scheinerstr.~1,
D-81679 M\"unchen,
Germany\\
$^{2}$ \quad
{Max Planck Institut for Astrophysics},
D-85748 Garching,
Germany}

\corres{Correspondence: rhea@usm.lmu.de}

\abstract{Recent hydrodynamic cosmological simulations cover volumes
up to Gpc$^3$ and resolve halos across a wide range of masses and
environments, from massive galaxy clusters down to normal galaxies,
while following a large variety of physical processes (star-formation,
chemical enrichment, {AGN}~feedback) to allow a self-consistent
comparison to observations at multiple wavelengths.
Using~the Magneticum simulations, we investigate the buildup of the
diffuse stellar component~(DSC) around massive galaxies within group
and cluster environments.
The DSC in our simulations reproduces the spatial distribution of the
observed intracluster light~(ICL) as well as its kinematic properties
remarkably well.
For galaxy clusters and groups we find that, although the DSC
in~almost all cases shows a clear separation from the brightest
cluster galaxy~(BCG) with regard to its \textit{dynamic} state,
the radial stellar \textit{density} distribution in many halos is
often characterized by a single S{\'e}rsic profile, representing both
the BCG component and the DSC, very much in agreement with current
observational results.
Interestingly, even in those halos that clearly show two components
in both the dynamics and the spatial distribution of the stellar
component, no correlation between them~is~evident.
}


\begin{document}

\section{Introduction}
Brightest cluster galaxies (BCGs), residing in the centers of galaxy
clusters, are the most massive and luminous galaxies in the universe.
During their lifetime, they experience frequent interactions with
satellite galaxies, and their growth is dominated by merger events.
These merger events also lead to the buildup of a diffuse stellar
component (DSC), which very likely contains a significant fraction
of the total stellar mass of the galaxy clusters (see
Murante et al., 2007~\cite{murante:2007} and references therein).
The~velocities~of the stars in the BCG and the DSC have distinct
kinematic distributions, which~can~be characterized by two
superposed Maxwellian distributions, as demonstrated by
Dolag et al., 2010~\citep{dolag:2010}.
While the velocity dispersion of the stars in the BCG represents
the central mass of the stars, the~velocity dispersion of the
DSC is much larger and is comparable to that of the dark matter
halo (see, for example,
Dolag et al., 2010~\cite{dolag:2010},
Bender et al., 2015~\cite{bender:2015}, and
Longobardi et al., 2015~\cite{longobardi:2015}).
More details on this matter can also be found in a recent review by
Mihos et al., 2016~\cite{mihos:2016}.

\textls[-25]{{Similarly, early simulations of galaxy clusters found
that the density distributions of BCGs in \mbox{clusters} can be
described by a superposition of two extended components as well (e.g.,
Puchwein et al., 2010~\cite{puchwein:2010}).}}
However, more recent simulations find the opposite, namely that in
many cases the radial density profiles can be described by a single
profile, which is in good agreement with observations.
These~\mbox{simulations} also indicate that a double-component fit
to the radial density profiles is only needed in rare cases.
Interestingly, the three-dimensional distribution of these outer
stellar halos seems to be described universally by a so-called Einasto
profile over a wide range of halo masses, as shown by
Remus et al., 2016~\cite{remus:2016}, where the curvature of the
radial profiles appears to be more closely linked to the cluster's
assembly history than the separation of the radial profiles into
distinct components.

In this study we analyse the velocity distributions as well as the
projected radial surface density profiles of the stellar component
in galaxy clusters selected from a state-of-the-art cosmological
simulation, and test for possible correlations between these
distributions.

\section{Simulations}
We use galaxy clusters selected from the
{Magneticum Pathfinder}
(\url{www.magneticum.org})
simulation set.
This suite of fully hydrodynamic cosmological simulations
comprises a broad range of~simulated volumes, with box lengths
of $2688~\mathrm{Mpc}/h$ to $18~\mathrm{Mpc}/h$, covering
different resolution levels of stellar particle masses from
\mbox{$m_\mathrm{Star}=6.5\times 10^{8}~M_\odot/h$}
at the lowest resolution level down to~particle masses of
\mbox{$m_\mathrm{Star}=1.9\times 10^{6}~M_\odot/h$}
at the highest resolution level.
For this work, we use two different simulations, Box2b and Box4,
with the smaller one (Box4) having a higher resolution.
The~details of these two simulations are summarized in
Table~\ref{tab:magneticum}.

\begin{table}[H]
  \caption{Magneticum simulations used in this work.}
  \label{tab:magneticum}
  \centering
  \begin{tabular}{ccccc}
    \toprule
    &
    \textbf{Box Size} &
    {\boldmath $N_\mathrm{part}$} & 
    {\boldmath $m_\mathrm{Star}$} &
    {\boldmath $\epsilon_\mathrm{Star}$} \\
    \midrule
    \textbf{Box2b hr} & 910~Mpc & $2\times2880^3$ & $3.5\times10^7~M_\odot/h$ & $2~\mathrm{kpc}/h$   \\
    \textbf{Box4 uhr} & 68~Mpc  & $2\times576^3$  & $1.9\times10^6~M_\odot/h$ & $0.7~\mathrm{kpc}/h$ \\
    \bottomrule
  \end{tabular}
\end{table}

All simulations of the Magneticum Pathfinder simulation suite are
performed with an advanced version of the tree-SPH code P-Gadget3
({Springel, 2005}~\cite{springel:2005}).
They include metal-dependent radiative cooling, heating from a uniform
time-dependent ultraviolet background, star formation according to
Springel \& Hernquist, 2003~\cite{springel:2003},
and the chemo-energetic evolution of the stellar population as traced
by {SN~Ia}, {SN~II}, and {AGB} stars, including the associated feedback
from these stars
\mbox{(Tornatore et al., 2007~\cite{tornatore:2007}).}
Additionally, they follow the formation and evolution of supermassive
black holes, including their associated quasar and radio-mode feedback.
For a detailed description, see
Dolag et al.~(in prep),
Hirschmann et al., 2014~\cite{hirschmann:2014}, and
Teklu et al., 2015~\cite{teklu:2015}.

Galaxy clusters are chosen according to the total mass of a structure
as found by the baryonic {\small SUBFIND} algorithm (see
Dolag et al., 2009~\cite{dolag:2009}).
For the larger, less resolved volume (Box2b), we classify all
structures with masses of $M_\mathrm{tot} > 2\times10^{14}~M_\odot$
as clusters, independent of
their dynamical state, and~find~890 objects.
For the smaller volume (Box4), there are no massive galaxy clusters,
but the increased resolution enables us to utilize halos with
masses down
to~\mbox{\thickmuskip=4mu \medmuskip=2mu
$1\times10^{13}~M_\odot < M_\mathrm{tot} < 1\times10^{14}~M_\odot$},
and therefore allows us to add galaxy groups down to
the limit of massive field galaxies to this study.
Including the three clusters and 35 groups from the smaller volume
simulation, we end up with~a~total sample of 928 objects, which
is an unprecedentedly large sample of simulated galaxy clusters
and~groups for which we here, for the first time, provide a
statistically representative analysis of the decomposition of the
stellar components into the BCG and the DSC, providing predictions
for future observational studies of the {ICL} and the BCGs.

\section{Velocity Distributions and Radial Surface Density Profiles}

\begin{figure}[p]
  \centering
  \colorbox{black}{
  \parbox{0.75\textwidth}{\includegraphics[width=\hsize]{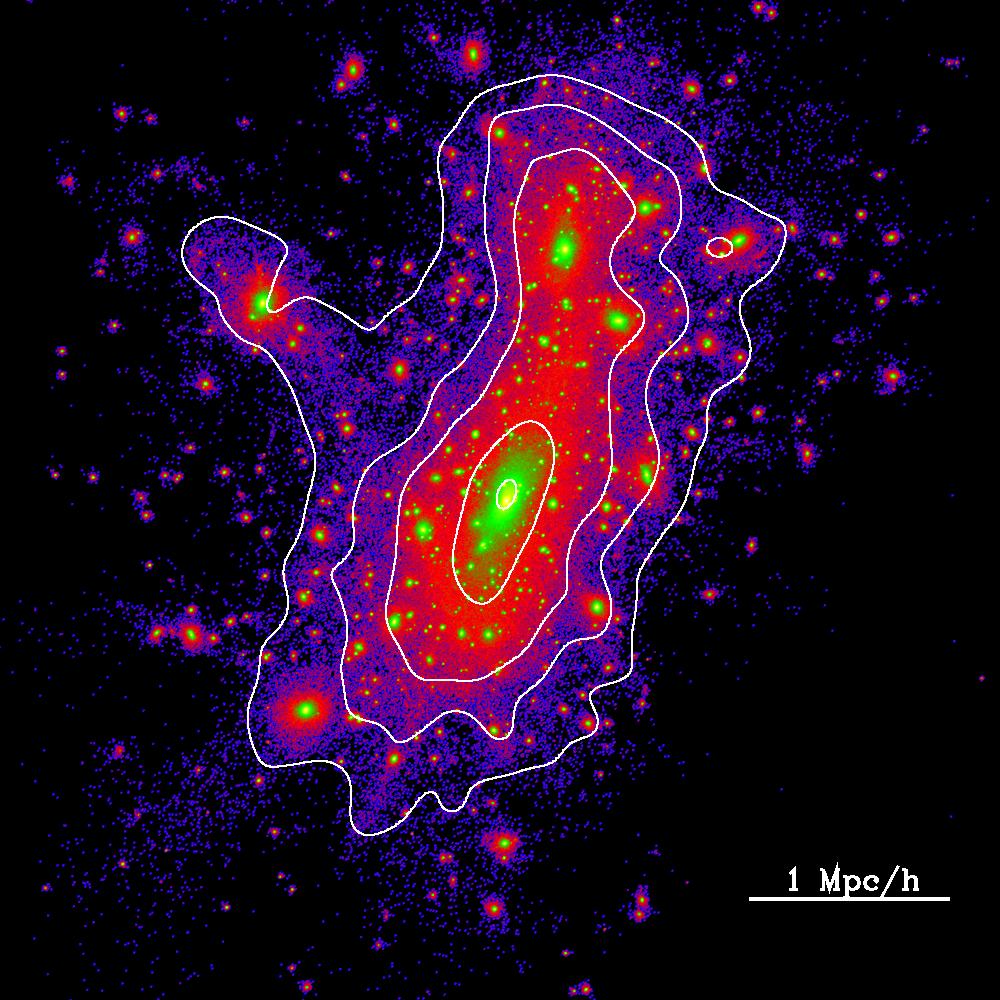}}%
  \parbox{0.2485\textwidth}{\includegraphics[width=\hsize]{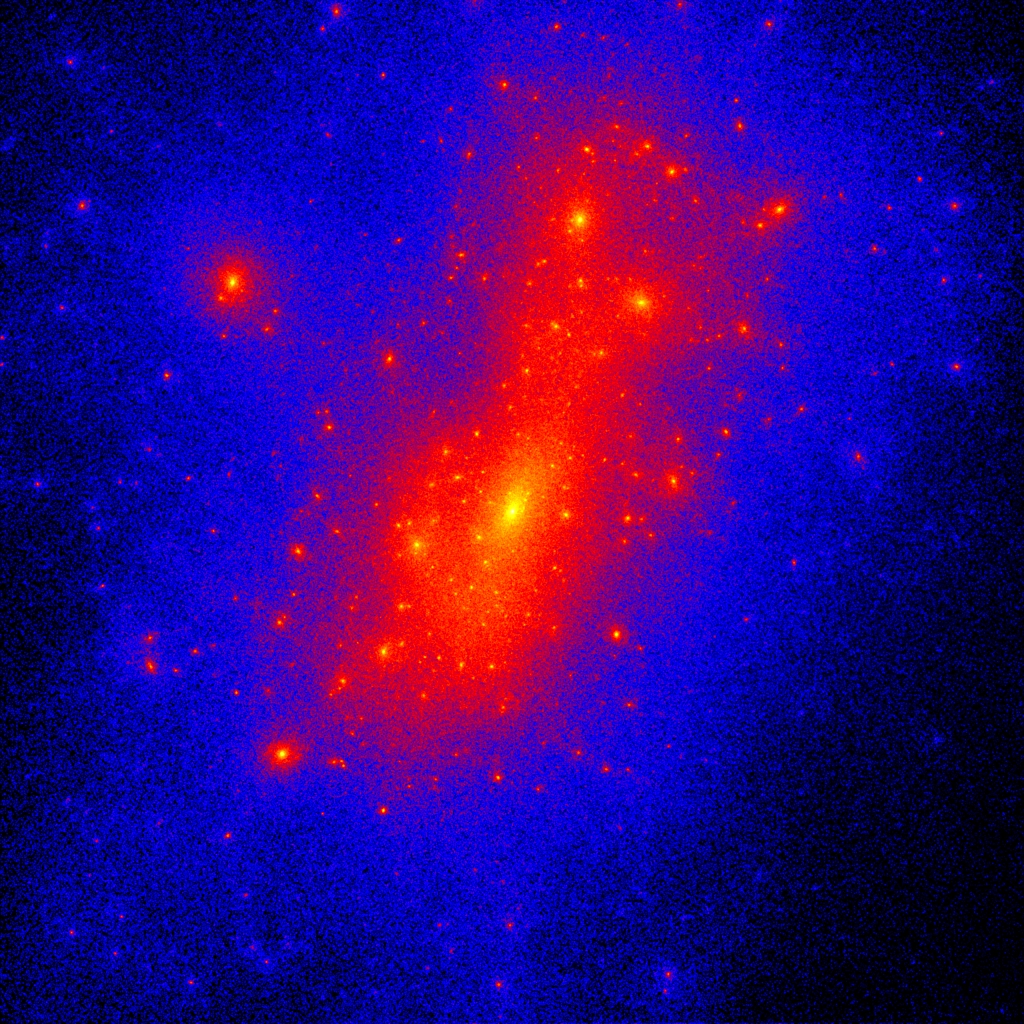}\\
                            \includegraphics[width=\hsize]{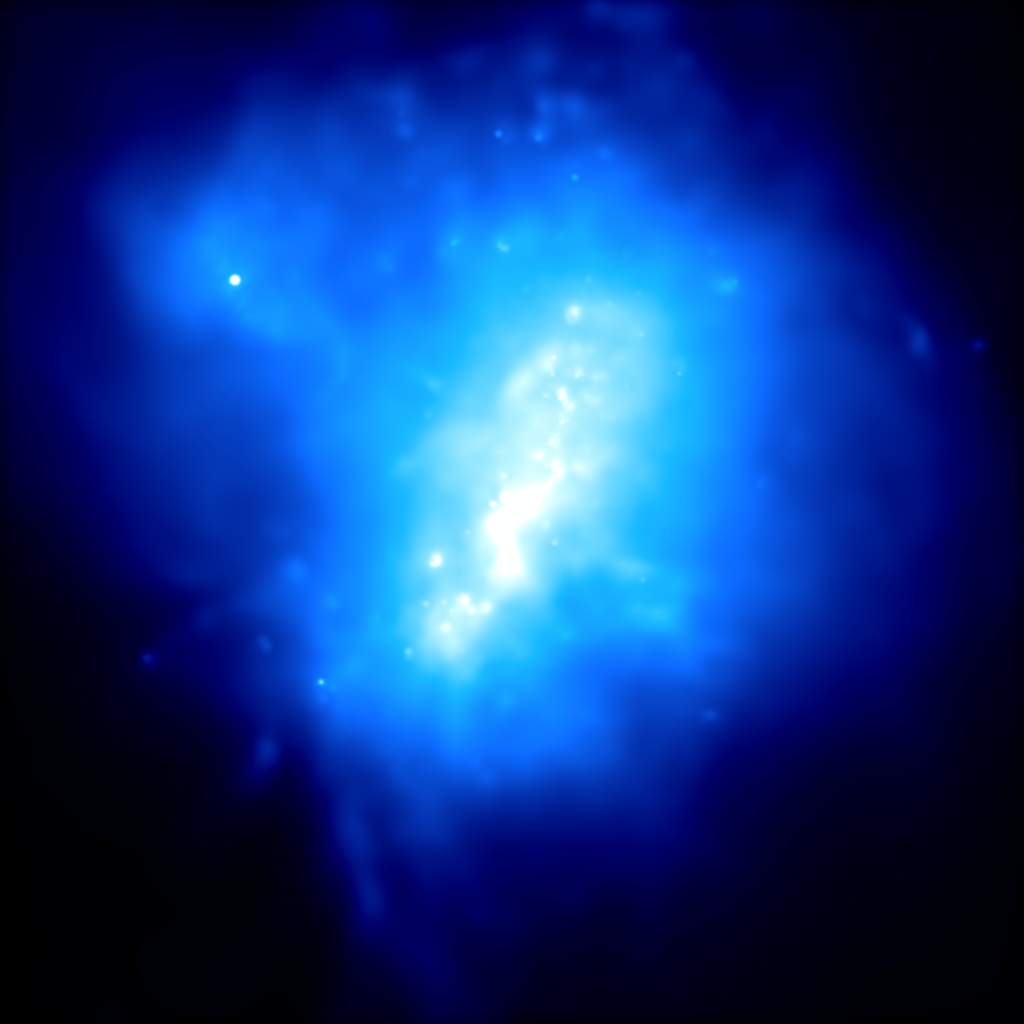}\\
                            \includegraphics[width=\hsize]{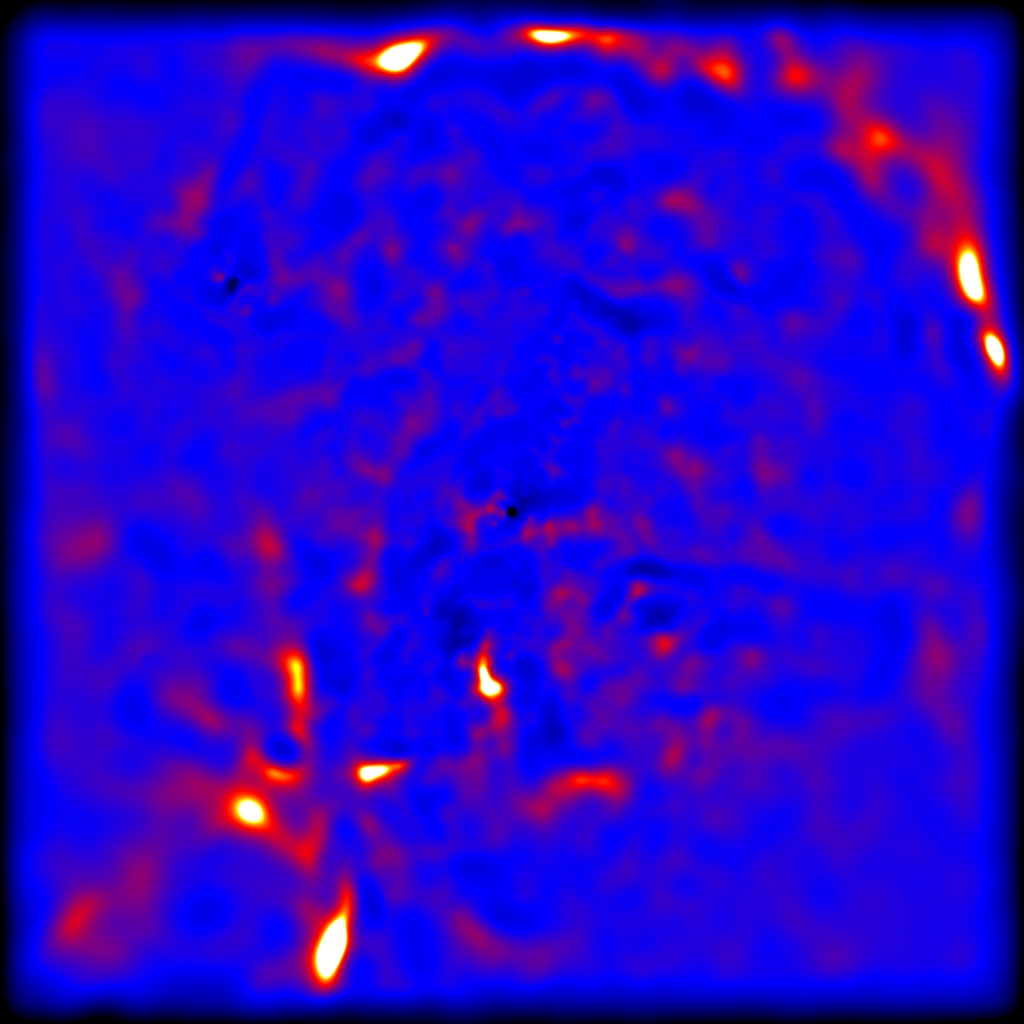}}}
  \includegraphics[width=0.95\textwidth]{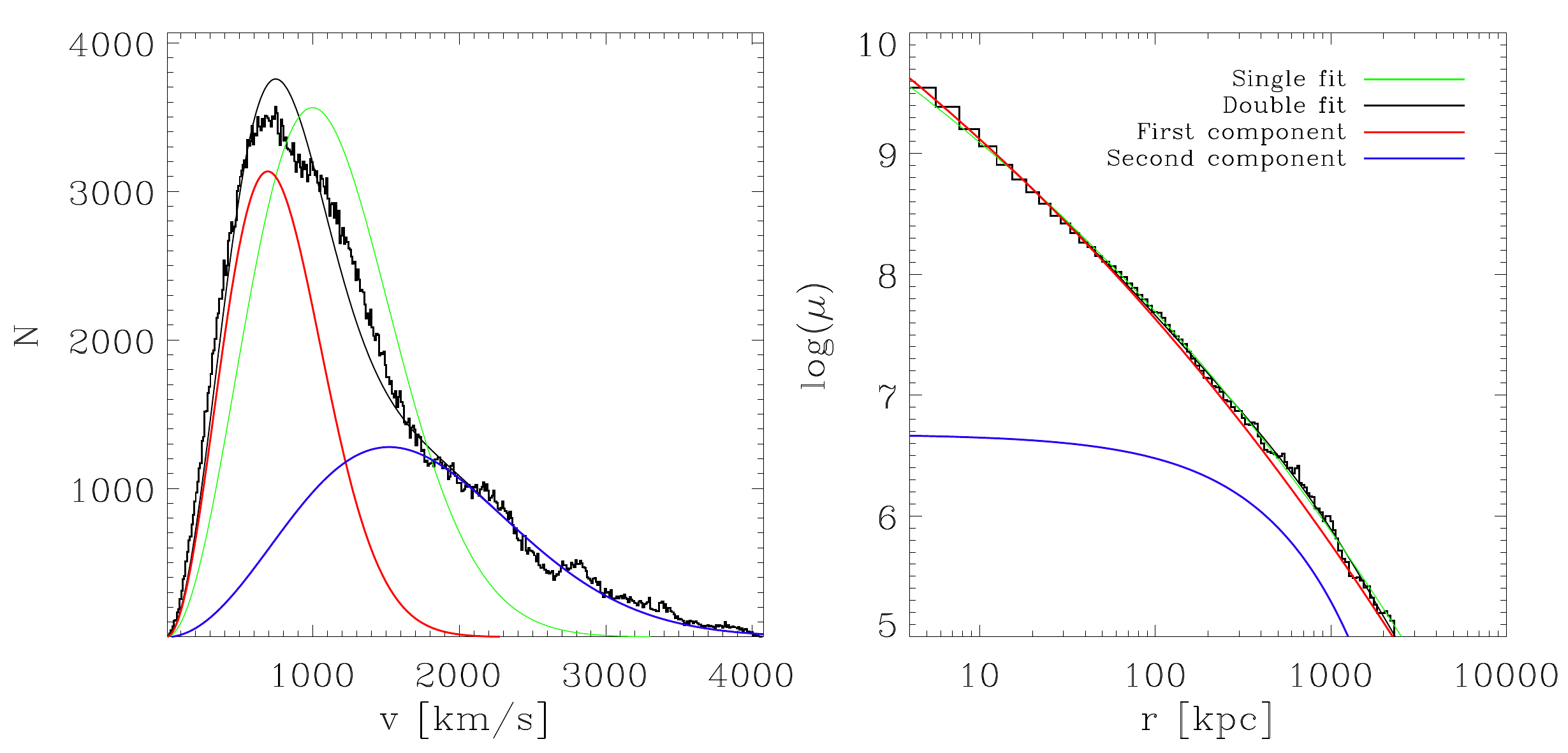}
  \caption{{Example of a galaxy cluster where the velocity distribution}
  is best described by a double Maxwellian fit, while the radial surface
  density profile can be well described by a single S{\'e}rsic profile
  (class d/s).
  {\textbf{Upper left panel}:}
  Stellar particle density map of the cluster, with the densest areas in
  yellow/green and the least dense areas in blue/black.
  White contours mark the iso-brightness lines of the {DSC} with the
  galaxies subtracted.
  {\textbf{Upper right panels, from top to bottom:}}
  Total matter density map; X-ray surface brightness map; unsharp-masked
  image of the pressure map to visualize shock fronts, as indicated by
  the large, arc-like feature in the upper right corner.
  {\textbf{Lower left panel:}}
  Velocity histogram for the stellar particles within the cluster,
  excluding those from substructures.\\
  \strut\hfill\textit{(Continued in Figure~\ref{fig:doubmax_doubsers}.)}
  }
  {\label{fig:doubmax_singsers}}
\end{figure}

\begin{figure}[p]
  \centering
  \colorbox{black}{
  \parbox{0.75\textwidth}{\includegraphics[width=\hsize]{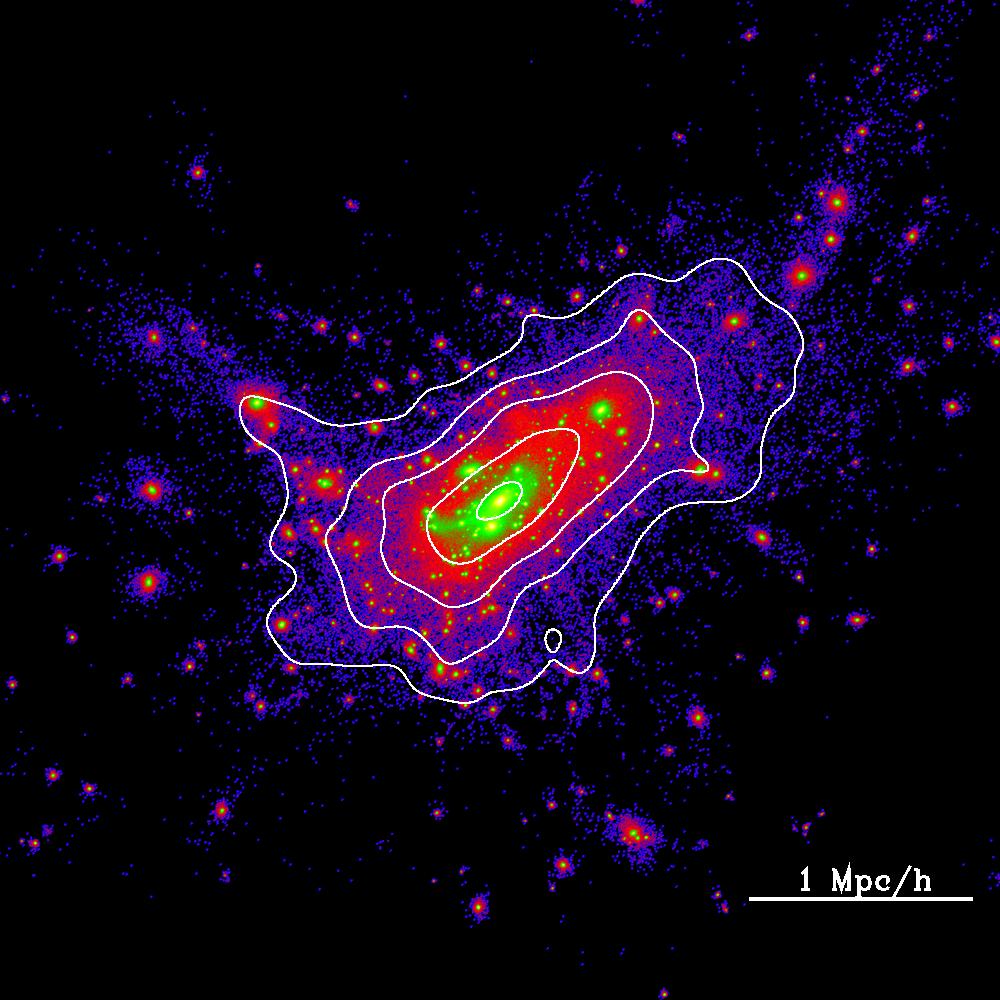}}%
  \parbox{0.2485\textwidth}{\includegraphics[width=\hsize]{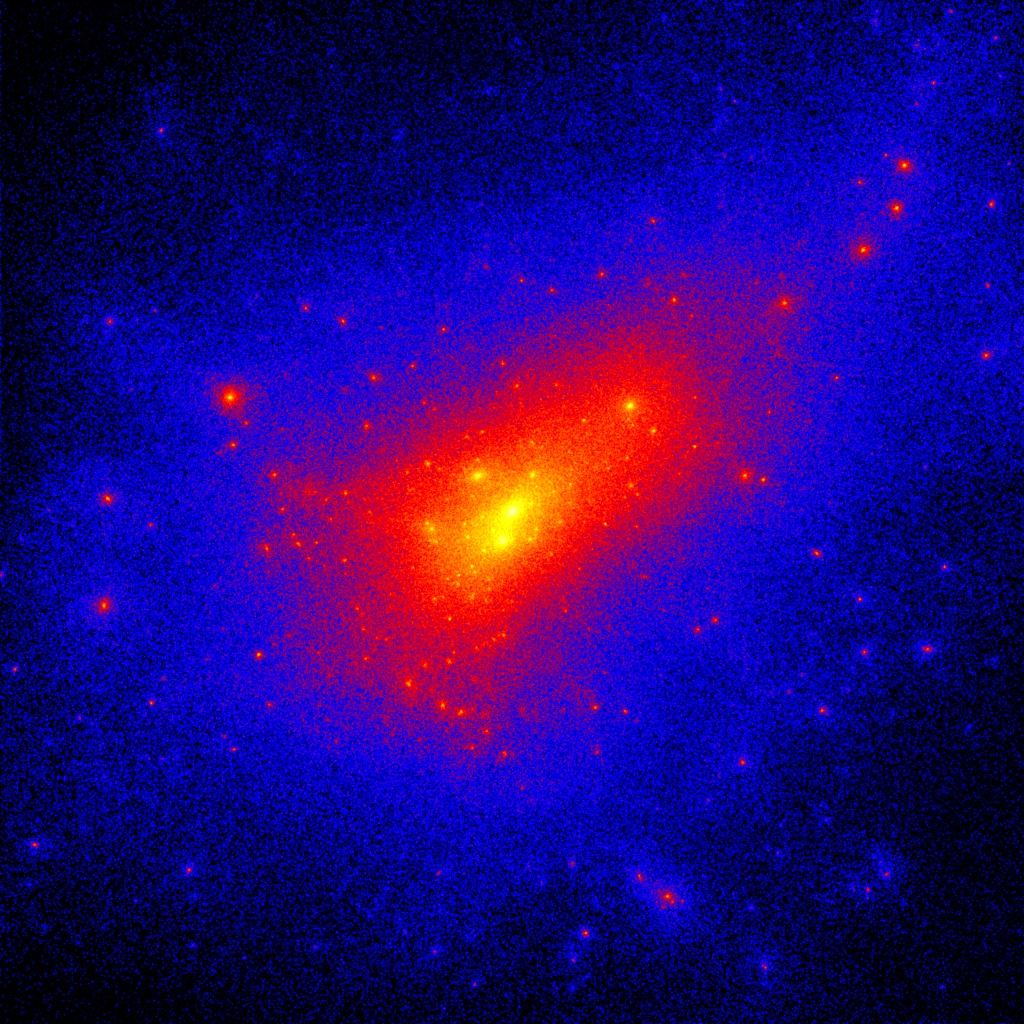}\\
                            \includegraphics[width=\hsize]{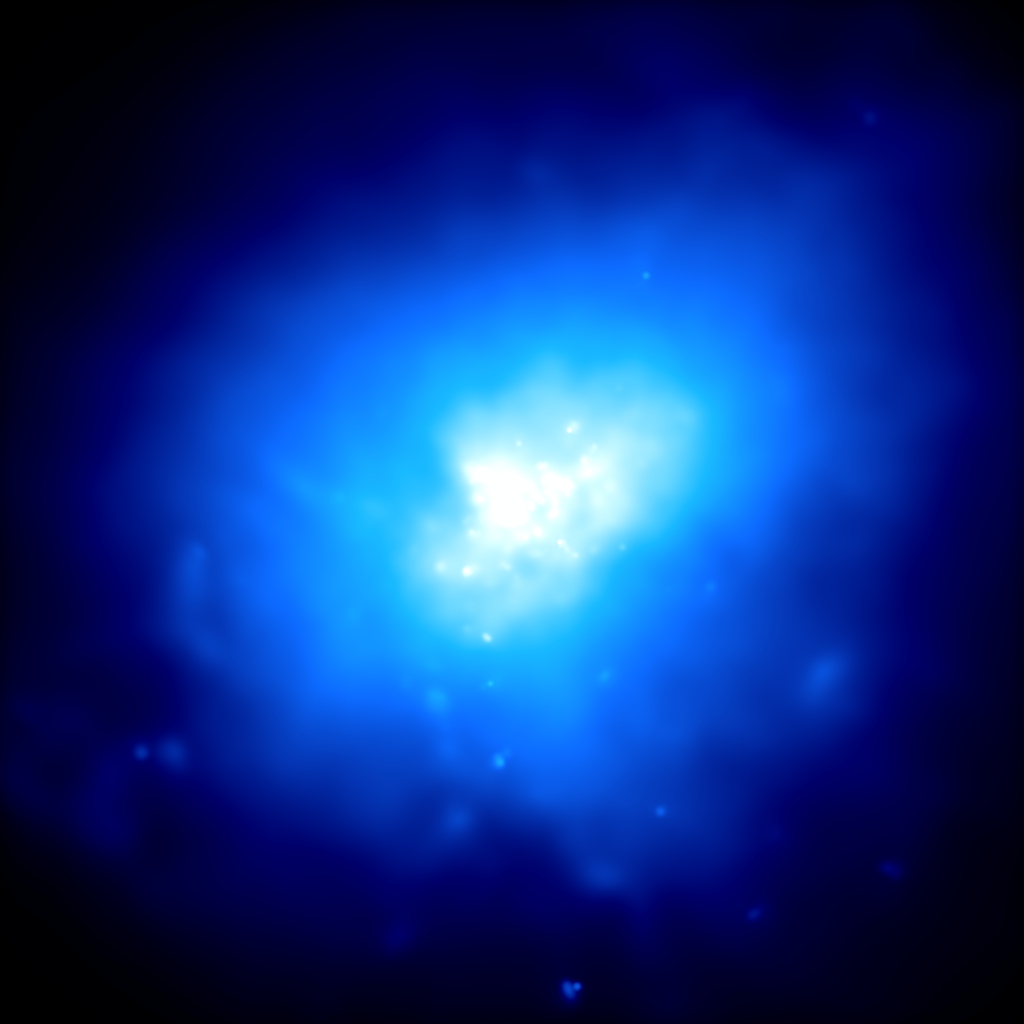}\\
                            \includegraphics[width=\hsize]{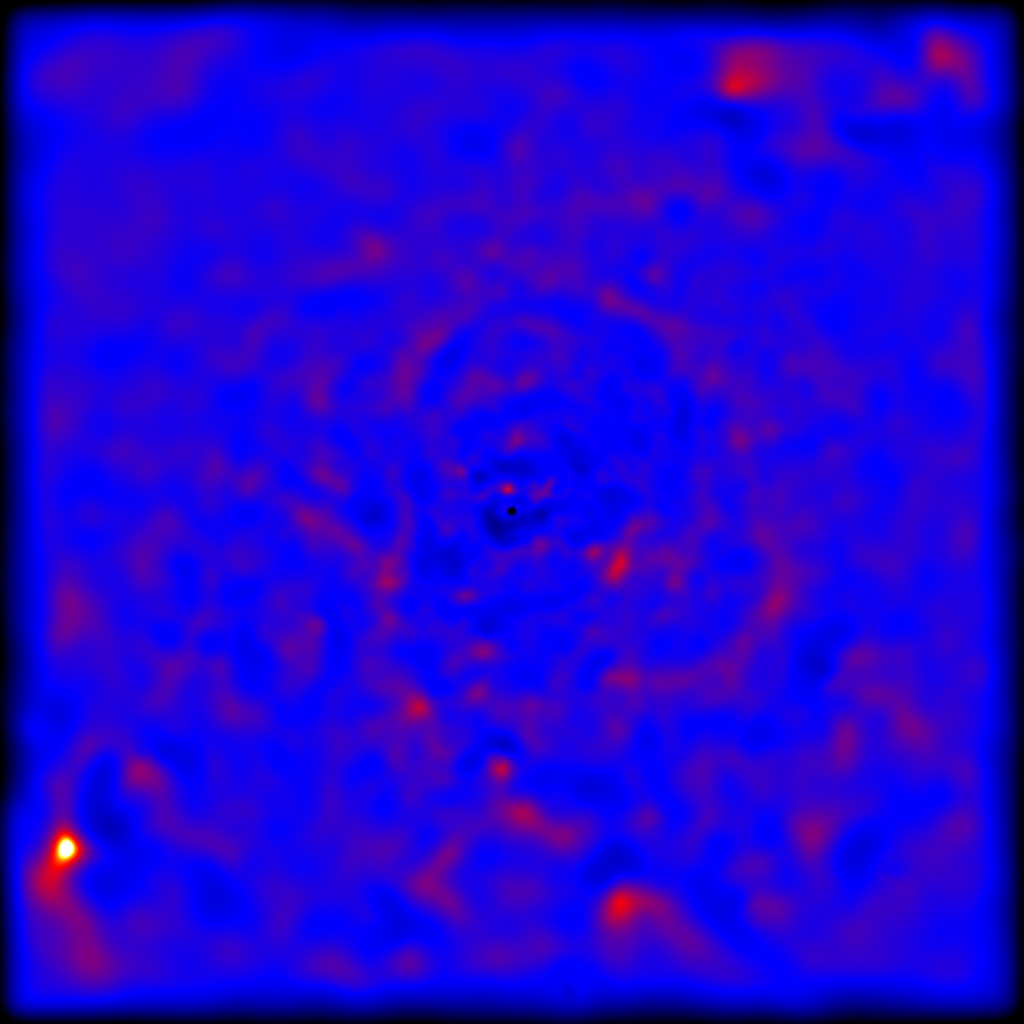}}}
  \includegraphics[width=0.95\textwidth]{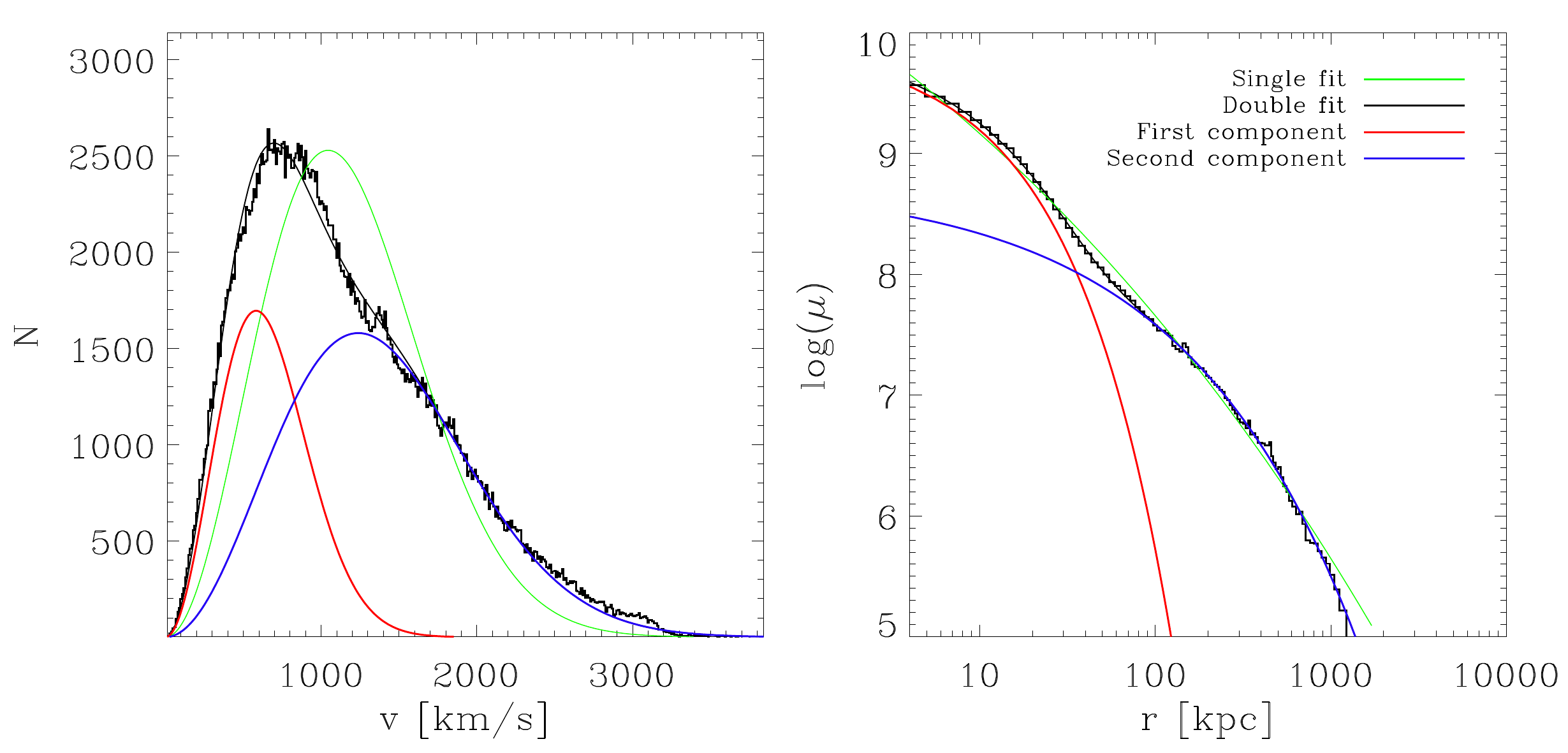}
  \caption{Same as Figure~\ref{fig:doubmax_singsers} {but for a}
  cluster with a double Maxwellian distribution in the velocity
  and~a~double S{\'e}rsic radial surface density profile as
  best representation (class d/d).
  Although the cluster is quite extended, there are no shocks visible,
  indicating a very late state of the merger.\\[\baselineskip]
  \textit{(Continued from Figure~\ref{fig:doubmax_singsers}.)}\quad
  The green line shows the best single-Maxwell fit to the histogram,
  while the black line shows the best double-Maxwell fit to the histogram
  with the red and blue lines indicating~the individual Maxwellians of
  the {BCG} (first) component and the DSC (second) component.
  {\textbf{Lower right panel:}}
  Projected radial stellar surface density profile (with substructures
  already subtracted) of the cluster centered around the BCG---colours
  as in the left panel but for the S{\'e}rsic fits.
  }
  {\label{fig:doubmax_doubsers}}
\end{figure}

\begin{figure}[p]
  \centering
  \colorbox{black}{
  \parbox{0.75\textwidth}{\includegraphics[width=\hsize]{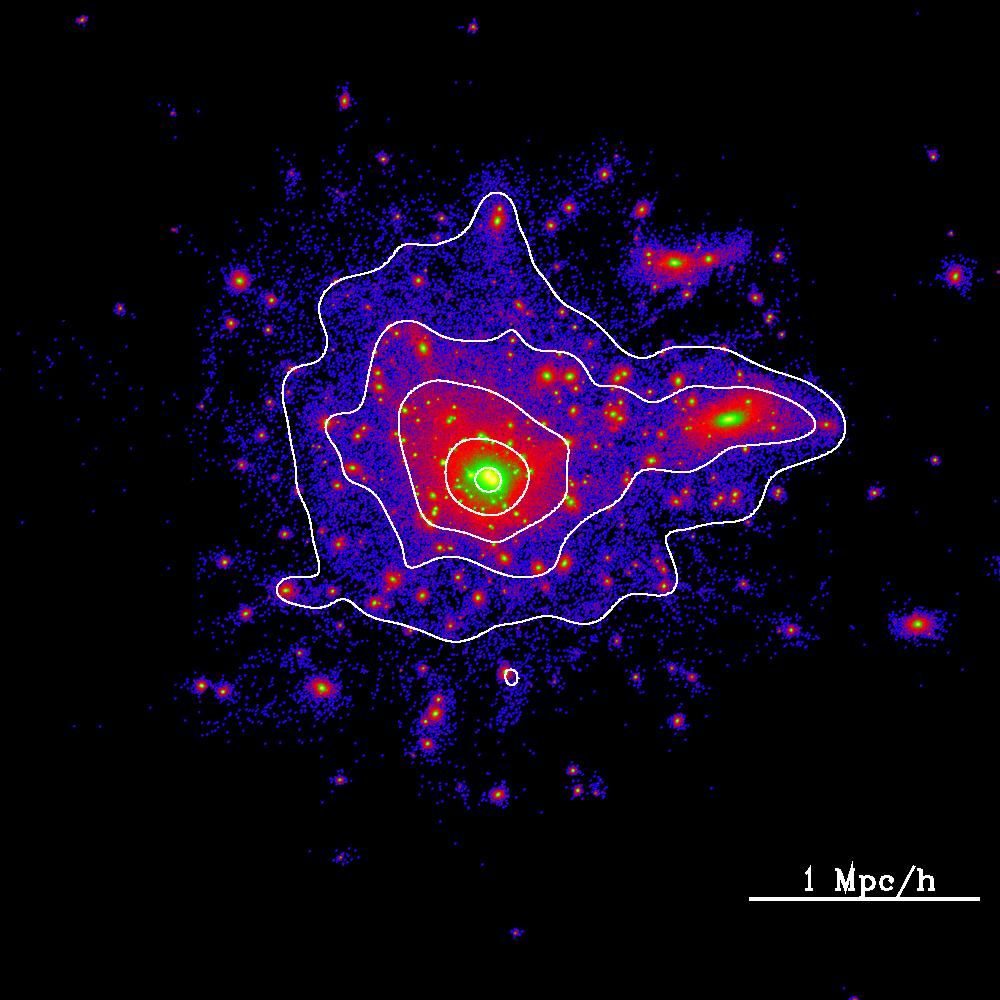}}%
  \parbox{0.2485\textwidth}{\includegraphics[width=\hsize]{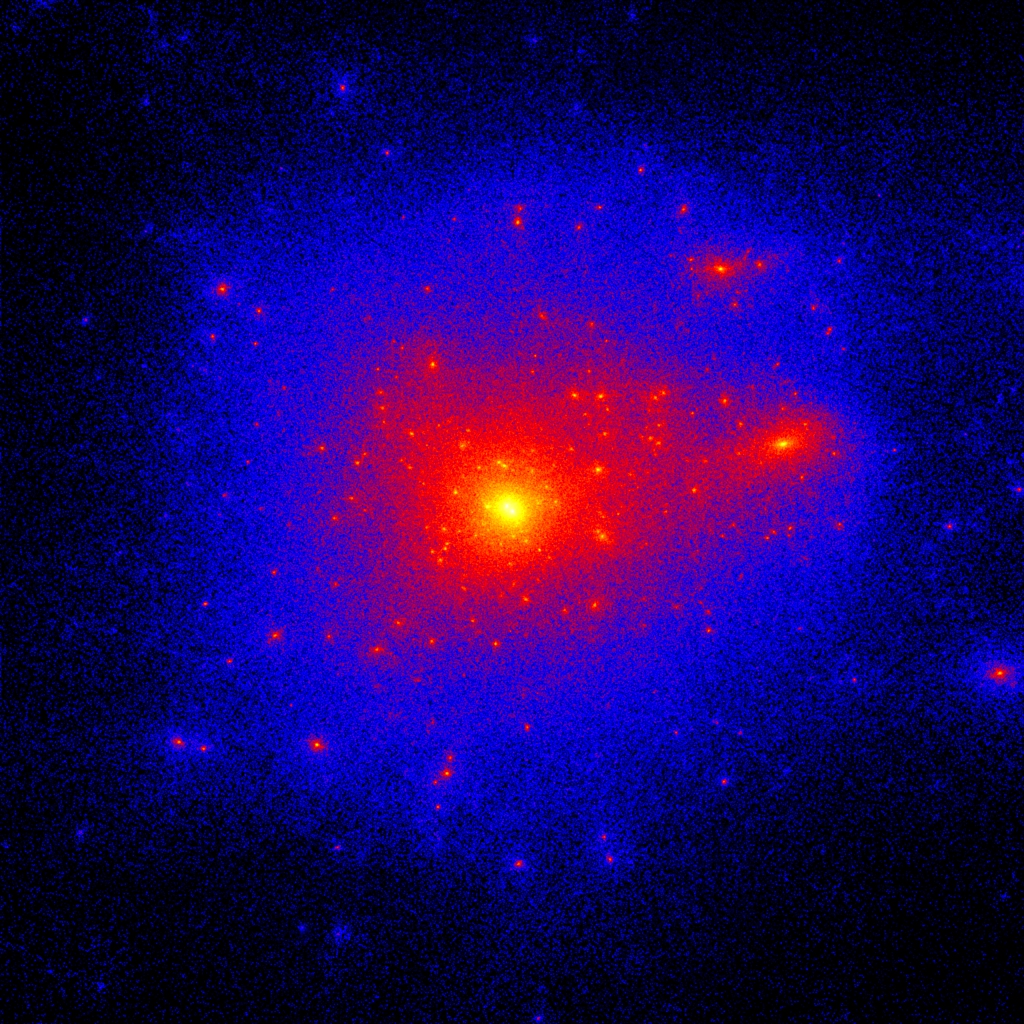}\\
                            \includegraphics[width=\hsize]{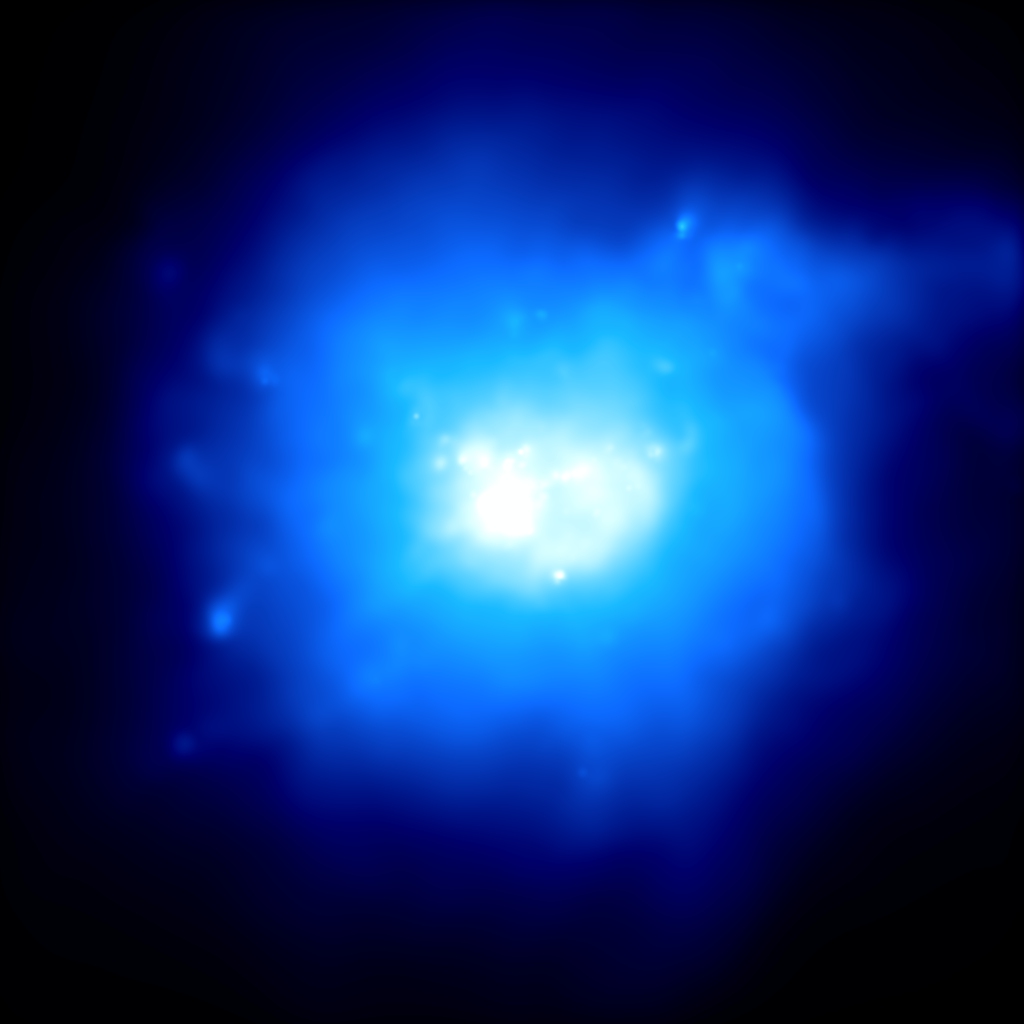}\\
                            \includegraphics[width=\hsize]{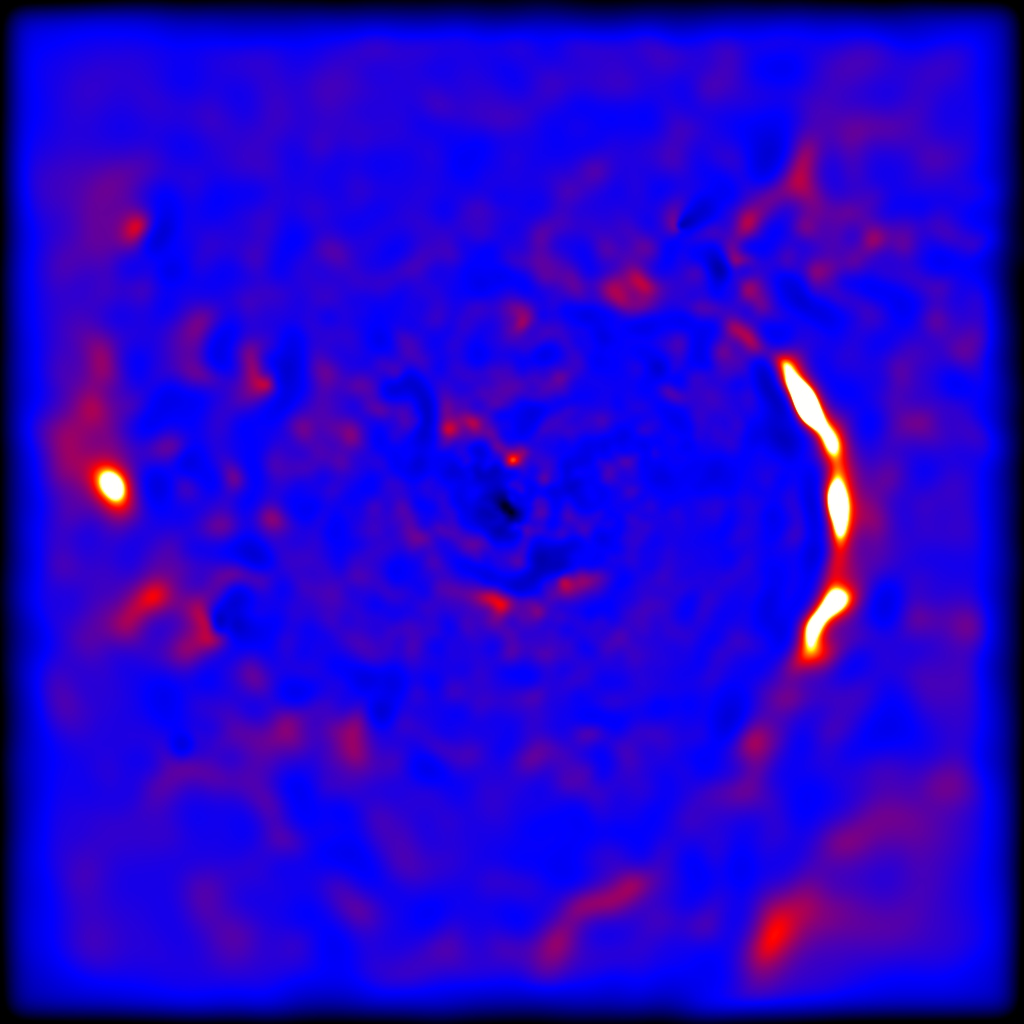}}}
  \includegraphics[width=0.95\textwidth]{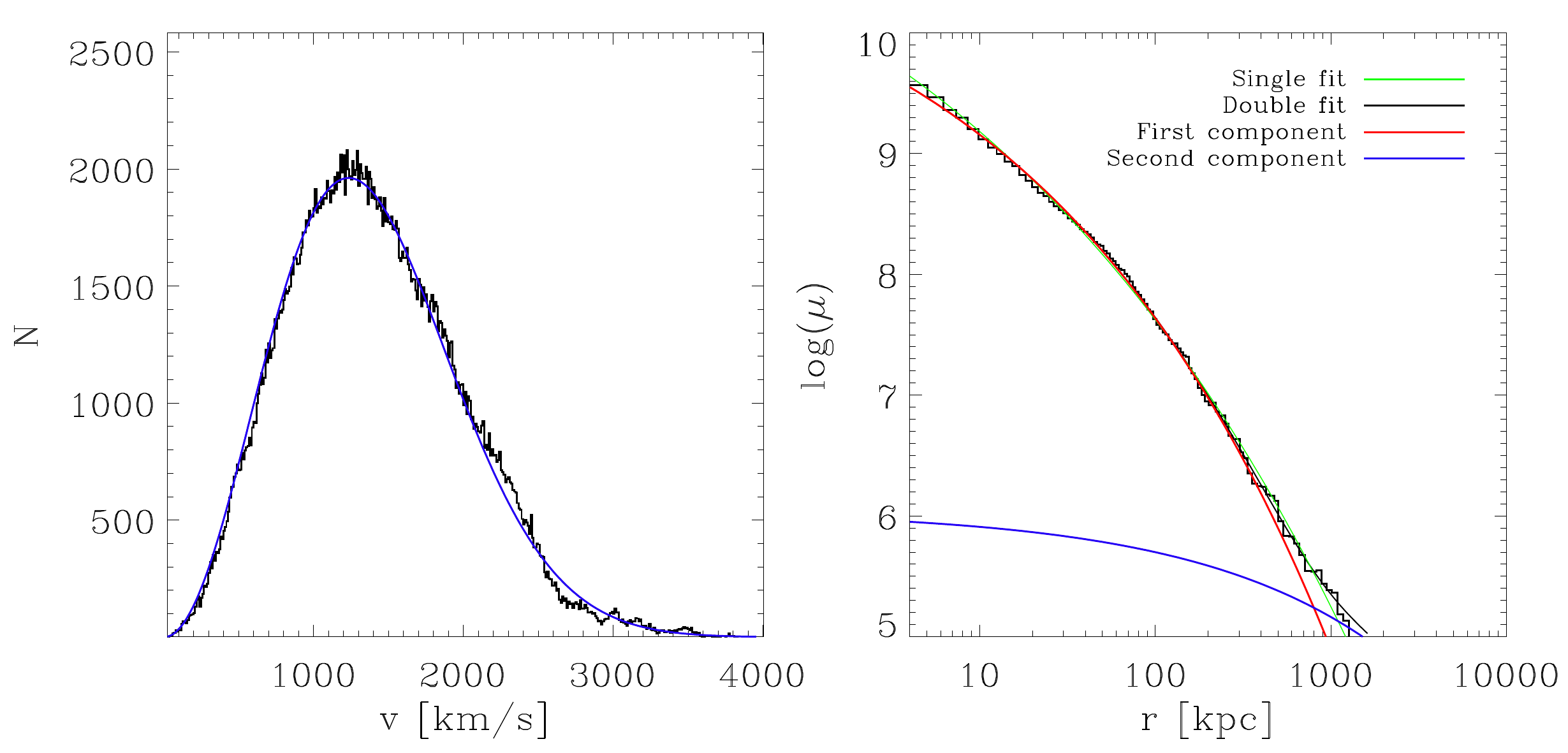}
  \caption{Same as Figure \ref{fig:doubmax_singsers} but for a cluster
  where the velocity distribution is well described by~a~single
  Maxwellian and the radial surface density profile is also well
  described by a single S{\'e}rsic profile (class s/s).
  Although the outwards-moving substructure to the right has lost all
  its gas component, the~injected shock within the {ICM} is still clearly
  visible as a large arc.
  }
  {\label{fig:singmax_singsers}}
\end{figure}

\begin{figure}[p]
  \centering
  \colorbox{black}{
  \parbox{0.75\textwidth}{\includegraphics[width=\hsize]{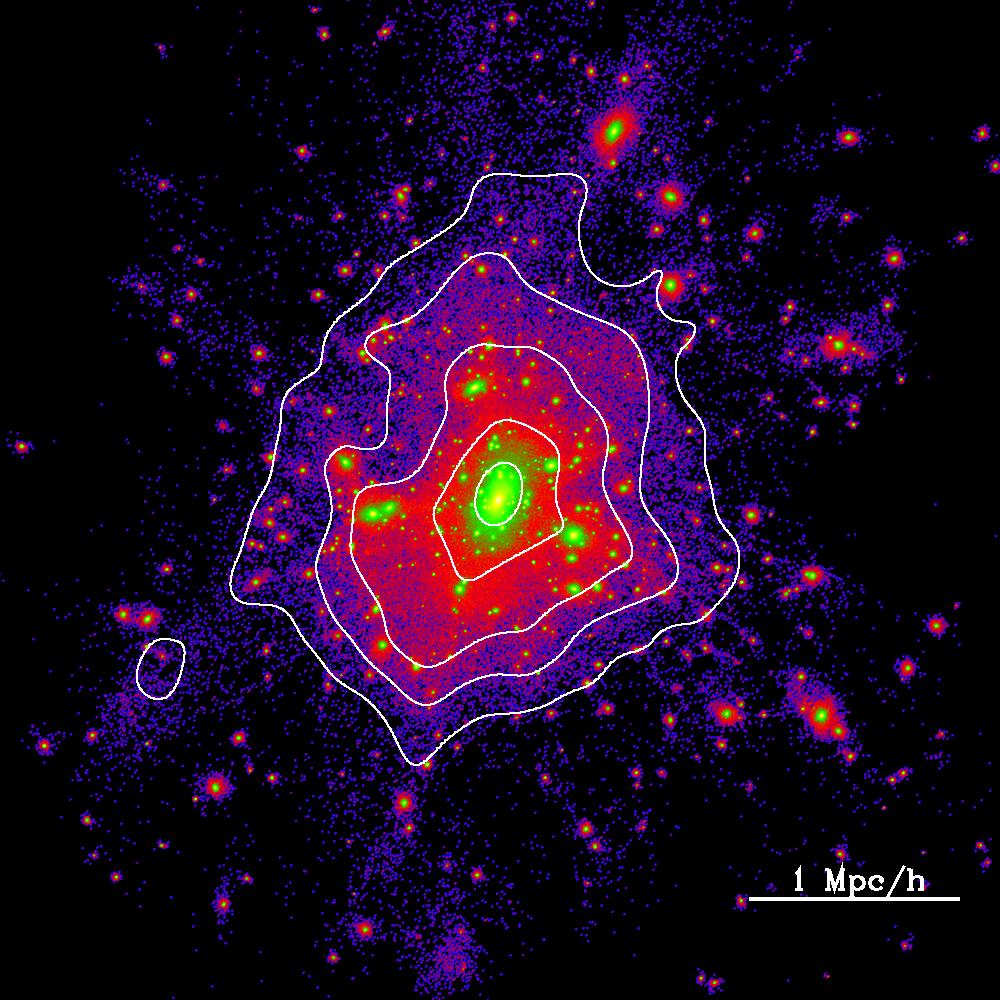}}%
  \parbox{0.2485\textwidth}{\includegraphics[width=\hsize]{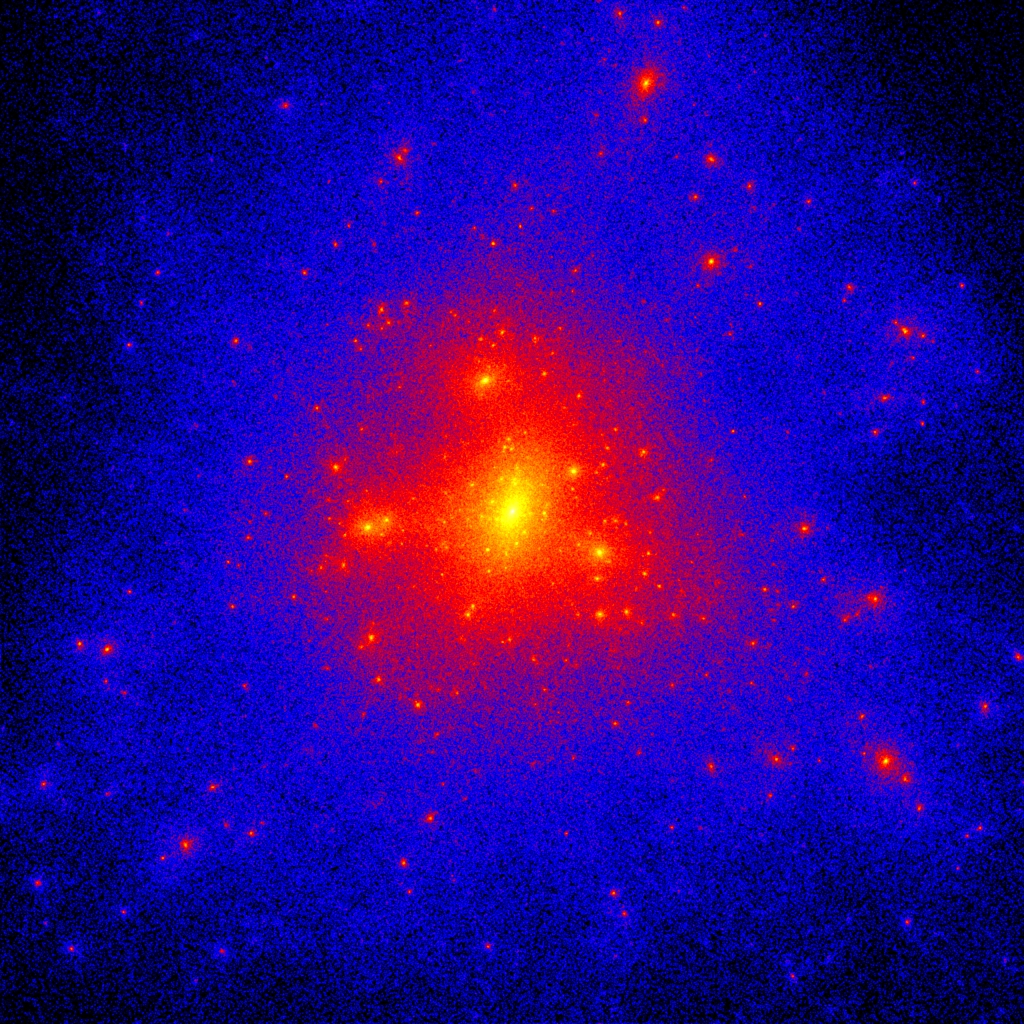}\\
                            \includegraphics[width=\hsize]{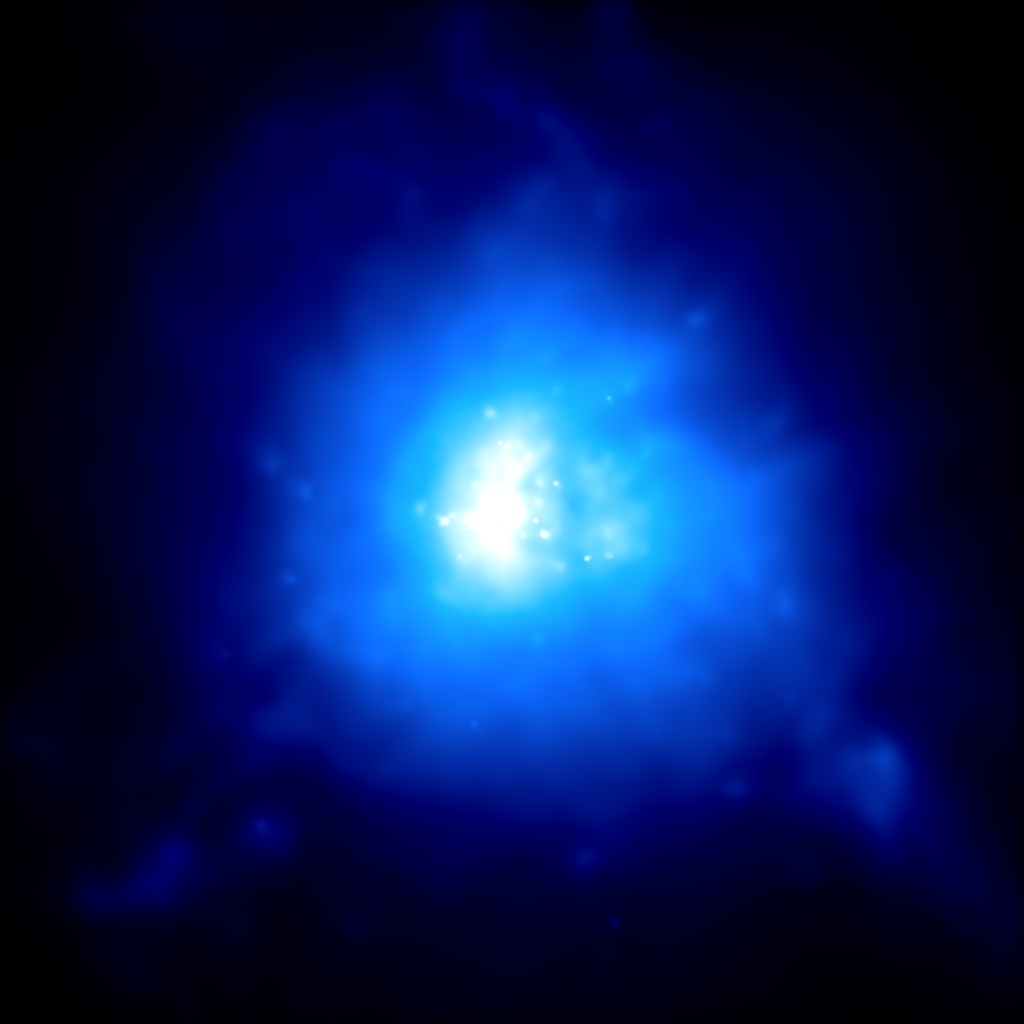}\\
                            \includegraphics[width=\hsize]{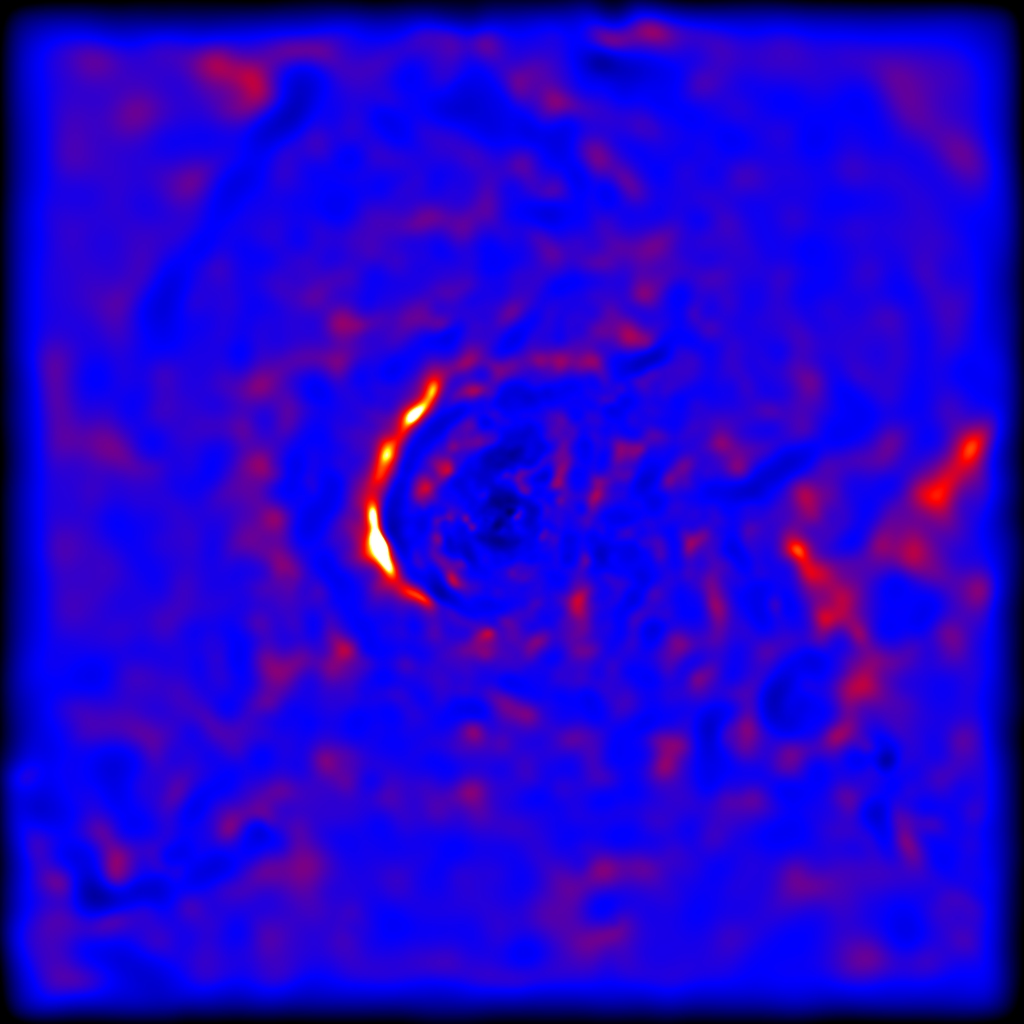}}}
  \includegraphics[width=0.95\textwidth]{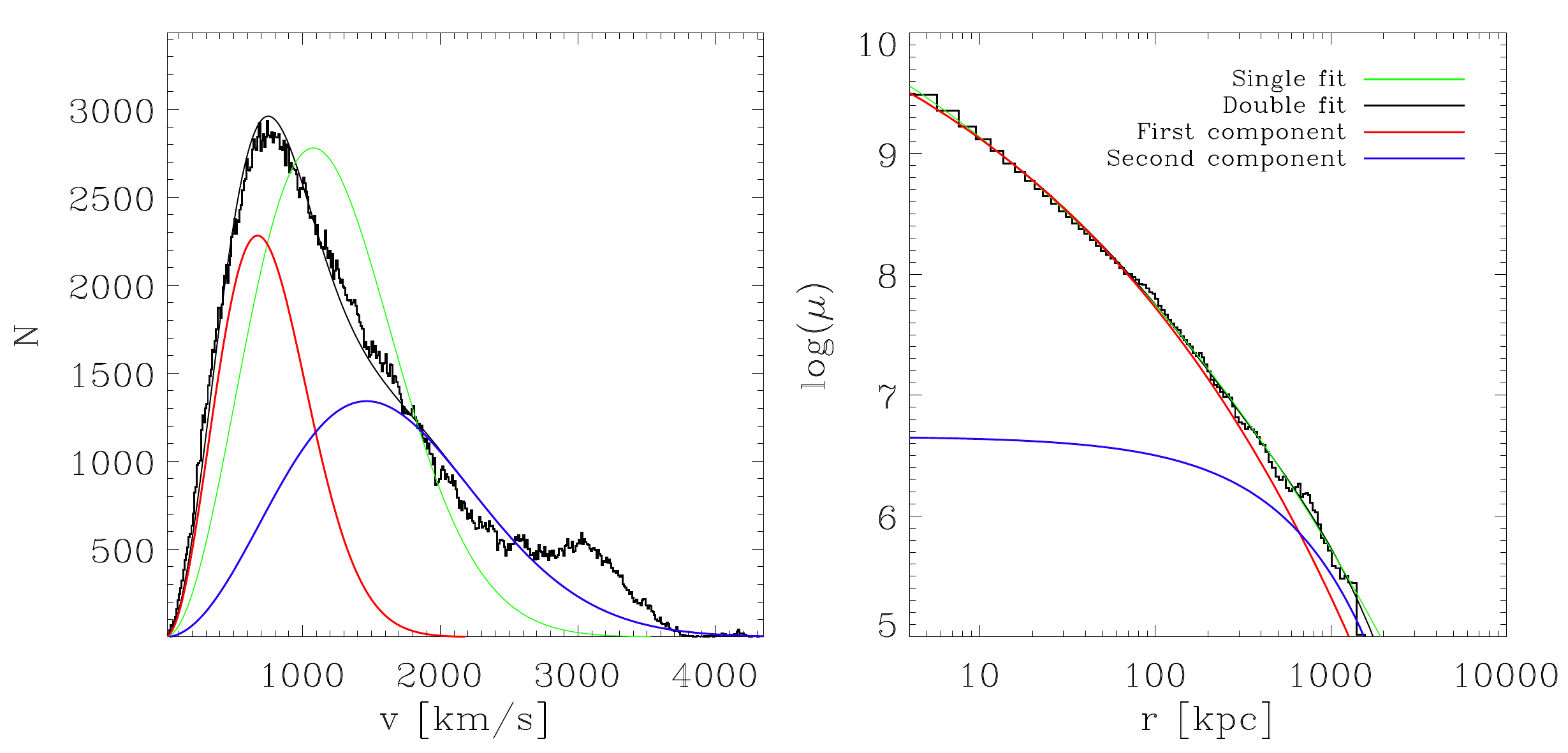}
  \caption{Same as Figure \ref{fig:doubmax_singsers} but for a special
  case where a third component would be needed to~describe
  the velocity distribution, while the radial surface density profile
  does not show signs of~a~third component.
  Since we do not explicitly study these special cases in this work, this
  cluster is classified as belonging belong to the d/d class, as both the
  velocity and the radial density clearly are multi-component systems.
  }
  {\label{fig:threemax_singsers}}
\end{figure}

In their detailed study, {Dolag et al., 2010}~\cite{dolag:2010}
demonstrated that the two dynamical components found in the velocity
distribution of the stellar component of galaxy clusters very well
represent the stellar component of the BCGs and the DSC, the latter
of which is itself a good approximation of the observed ICL in galaxy
clusters.
Following their approach, we subtract all substructures
(identified with {\small SUBFIND}) from the stellar component of each
cluster and use the remaining stars for this analysis.
First,~we calculate the velocities of all stellar particles in
a cluster and bin them in small equal-width bins of
$\Delta v = 10~\mathrm{km/s}$, thereby obtaining the intrinsic 3D
velocity distribution of the stars in each cluster.
Similarly, we choose a random viewing angle and calculate the projected
radius of each stellar particle.
Subsequently, we radially bin these particles using equal-particle
bins, thus obtaining radial surface density distributions, effectively
mimicking the radial surface brightness profiles that are commonly
observed for galaxies and galaxy clusters, assuming a constant
mass-to-light ratio.
Examples of the velocity distributions and surface density
profiles obtained by this methods are shown in the lower panels of
Figures~\ref{fig:doubmax_singsers}--\ref{fig:threemax_singsers}.


To obtain the different components of BCG and DSC in the simulations,
we again follow \mbox{Dolag et al., 2010~\cite{dolag:2010}}.
First, we fit a superposition of two Maxwellian distributions
\begin{equation}
N(v) = k_1v^2\exp\left(-\frac{v^2}{\sigma_1^2}\right)
     + k_2v^2\exp\left(-\frac{v^2}{\sigma_2^2}\right)
\end{equation}
{to the velocity distribution of each cluster.}
Additionally, we fit a single Maxwellian distribution to~the velocity
distributions for comparison purposes.
In most cases, a double-Maxwellian fit is needed to properly represent
the underlying velocity distributions, as shown, for example,
in the lower left panels of Figures~\ref{fig:doubmax_singsers}
and~\ref{fig:doubmax_doubsers}.
For comparison, we also show the stellar particle surface density
map of the clusters including all substructures in the large image
at the top of these figures.
The white contours show equal-density lines of the stellar distribution
without the substructures.
For both clusters shown in Figures~\ref{fig:doubmax_singsers}
and~\ref{fig:doubmax_doubsers}, the BCG is clearly visible, but while
the velocity distributions can be well described by double-Maxwellian
fits in both cases, the morphological appearance of the two clusters
is very different:
while the cluster shown in Figure~\ref{fig:doubmax_singsers} is clearly
elongated with a massive colliding structure clearly visible even
in the dark matter component (upper small image), the other cluster
shown in Figure~\ref{fig:doubmax_doubsers} shows no signs of ongoing
substantial accretion, and is only slightly elongated.
This is true even in the X-ray map (middle small image), where the
elongation is clearly visible for the first cluster while the second
cluster shows a more compact shape.
We also do not find a similarity between these clusters with regard
to their shock properties:
whereas the cluster shown in Figure~\ref{fig:doubmax_singsers}
has a clearly visible shock front in the upper right area of the
cluster, indicating a recent merging event, the cluster shown in
Figure~\ref{fig:doubmax_doubsers} shows no clear signs of such a
recent merger event in the shock map (bottom small image).
This clearly indicates that the velocity distribution of the cluster
remembers the merger history of the cluster over a much larger
timescale than other tracers like shocks or satellite distributions,
which provide information only about the more recent mass assembly
history of a cluster.

In some cases, there is no improvement to the description of
the velocity distribution of~an~individual cluster by using
a double-Maxwellian distribution for the fit, as the velocity
distributions of that particular cluster is already well described
by a single Maxwellian distribution (see,~for example,~the lower left
panel of Figure~\ref{fig:singmax_singsers}).
While the single-Maxwellian fit is a good approximation to the velocity
distribution of the cluster stellar light, the stellar light map in
the large image in the same figure clearly shows that the cluster is
currently accreting another, relatively massive, substructure.
This can be seen not only in the stellar component but also in the
X-ray emission (middle small image) and the shock map (bottom small
image).
Thus, this clearly shows that the contribution from this merger to
the cluster's DSC is not very large yet and does not show up in the
velocity distribution of the cluster as most stars that are brought
in through the merger event have been subtracted by {\small SUBFIND}.
Only~in the very-high velocity end of the velocity distribution,
the newly accreted component starts to be visible.
This also indicates that the DSC of this cluster is very rich, which
is caused by a rather diverse accretion history.
However, these cases are very rare as we will show later on in
this work.

\makeatletter
\setlength{\@fptop}{0pt}
\setlength{\@fpbot}{0pt plus 1fil}
\makeatother

The lower left panel of Figure~\ref{fig:threemax_singsers} shows a
very interesting albeit rare case for the velocity distribution of
a galaxy cluster:
for this cluster, a double-Maxwellian fit is still not sufficient and
a third superposed component would be needed to actually capture all
features visible in the velocity distribution.
Interestingly, this new accretion is not strongly visible in the maps
in the upper panels of this figure.
The~stellar component of the merger ongoing in the central part of the
cluster---clearly visible through the displacement of the central X-ray
emission from the center of mass towards the left as well as through
the clearly visible shock moving from the center leftwards---is most
likely the third, high velocity component in the velocity distribution.
As these cases are rare and need a visual inspection of all 928
clusters to identify them, we will not introduce these objects as a
separate class of~velocity distributions in this work, and will here
simply classify them as double-Maxwellian velocity distributions.

\begin{figure}[t]
  \centering
  \includegraphics[width=0.5\textwidth]{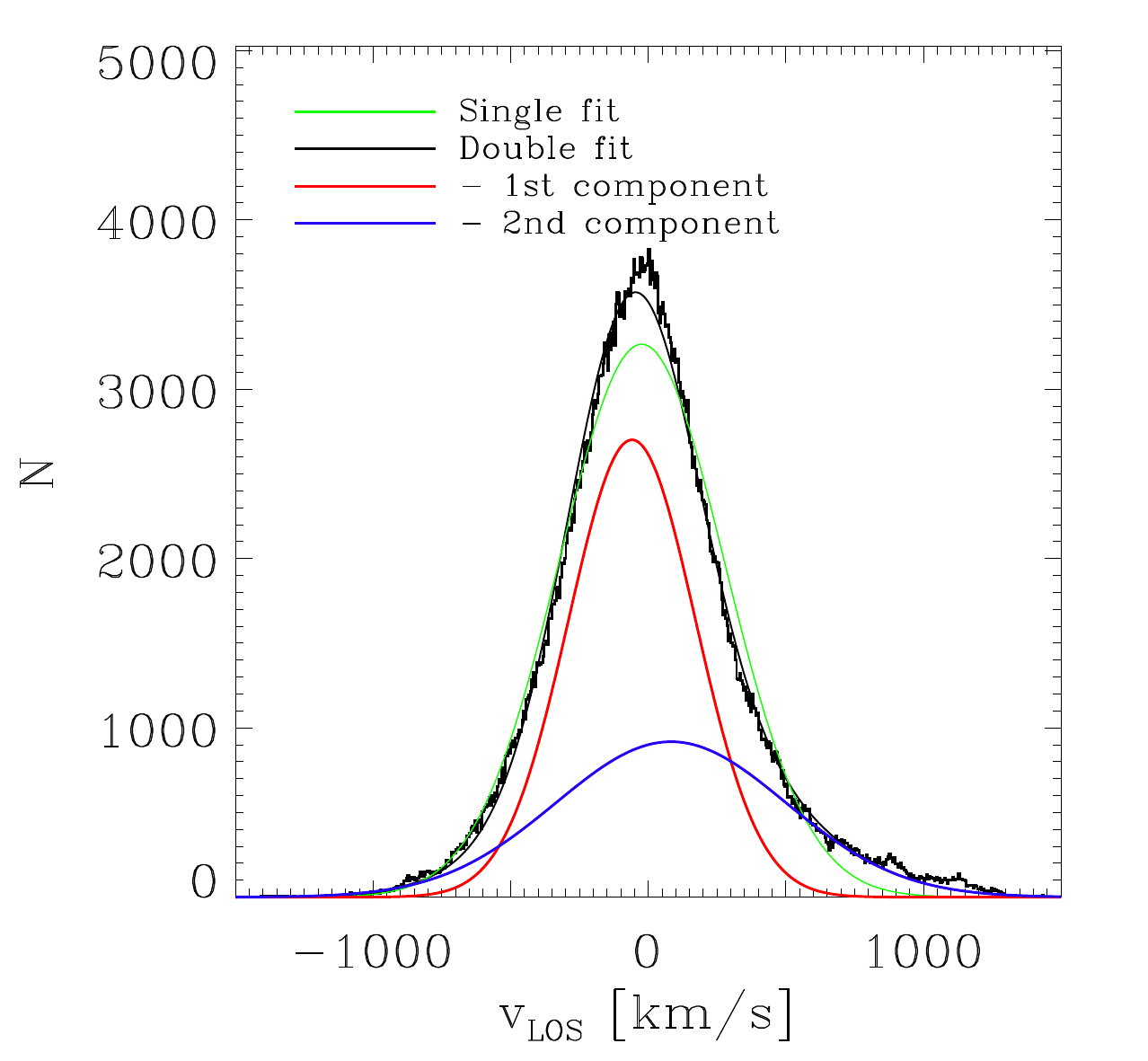}
  \caption{Line-of-sight velocity distributions of the stellar particles
  (excluding those bound to~substructures) of the cluster shown in
  Figure~\ref{fig:doubmax_singsers}, from a random projection.
  The green line is the best single Gaussian fit to the distribution,
  while the black line is the best double-Gaussian fit with its
  components shown in red and blue.}
  {\label{fig:gauss}}
\end{figure}

While in this three-dimensional analysis the two different velocity
components of the BCG and the DSC are nicely visible, this information
cannot be drawn directly from observations.
However,~observationally, the line-of-sight velocity can be measured
and used as a substitute to distinguish the two components:
In projection, the measured velocity is not represented
by~a~Maxwellian distribution but by a Gaussian distribution,
an example of which is shown in~Figure~\ref{fig:gauss}, where
we plot the line-of-sight velocity distribution of the cluster
\mbox{shown~in~Figure~\ref{fig:doubmax_singsers}.}
As for the three-dimensional case, also in the projected case a
superposition of two Gaussian fits is needed to represent the velocity
distribution profile, clearly indicating the two-component structure
of the BCG and the DSC.
In case of an ideal spherically symmetric relaxed system, the
projected Gaussian fits predict the same mass fractions and velocity
dispersions for the BCG component and the DSC as the intrinsic
Maxwellian distribution fits, but projection effects, asymmetries
as well as distortion effects through accretion events can lead to
(slightly) different values.
However, we do not investigate these issues further in this work.

For the radial surface density profiles, we use a similar approach
as for the velocity distributions.
We fit the superposition of two S{\'e}rsic profiles
\begin{equation}
\mu(r) = \mu_1 \exp\left(-\left(\frac{r}{r_1}\right)^{1/n_1}\right)
       + \mu_2 \exp\left(-\left(\frac{r}{r_2}\right)^{1/n_2}\right)
\end{equation}
{to the radial surface density profiles of each cluster, and, for
comparison, we also fit a single S{\'e}rsic~profile~ as well.}

Examples for the resulting S{\'e}rsic fits to the surface
density distributions are shown in the lower right panels of
Figures~\ref{fig:doubmax_singsers}--\ref{fig:threemax_singsers}.
As for the velocity distributions, we also see that there is no clear
correlation between the visual appearance of the clusters, neither in
the stellar nor the X-ray or the shock appearance, and the necessity
of a double-S{\'e}rsic fit.
Interestingly, we can already see from these four examples that
there is also no clear correlation between the presence of a second
component in the velocity distribution and the presence of a second
component in the radial density profile:
as~shown in Figure~\ref{fig:doubmax_singsers}, there are clusters
that display two components in the velocity distribution but
only a~single component in the radial surface density profile,
while Figure~\ref{fig:doubmax_doubsers} shows a cluster where both
distributions have a double structure.
In the following, we will study this behaviour in more detail.

\subsection{Statistical Properties}

As we cannot check the properties of all 928 galaxy clusters from
our sample individually, we~now try to quantify their behaviour in
a more statistical way.
The biggest issue here is that a double fit with twice as many free
parameters as a single fit will always yield a better fit, or one at
least as good, according to simple statistical tests like $\chi^2$
or Komolgorov--Smirnov, if the number of degrees of freedom is much
larger than the number of fit parameters (as is the case here).
Thus, we need to~find a better way to decide which fit adequately
characterizes the properties in velocity and surface brightness of
a cluster.

One way to do this is to use the double-Maxwell and double-S{\'e}rsic
fits of a cluster and integrate over each of the two components.
This way, assuming that the two components always represent an~inner,
slower component that describes the BCG and an outer, faster component
that describes the DSC, we can obtain the fraction of mass associated
with each component, relative to the total stellar mass given by the
full velocity distribution and the full radial surface density profile.

\begin{figure}[t]
  \centering
  \includegraphics[width=1\textwidth]{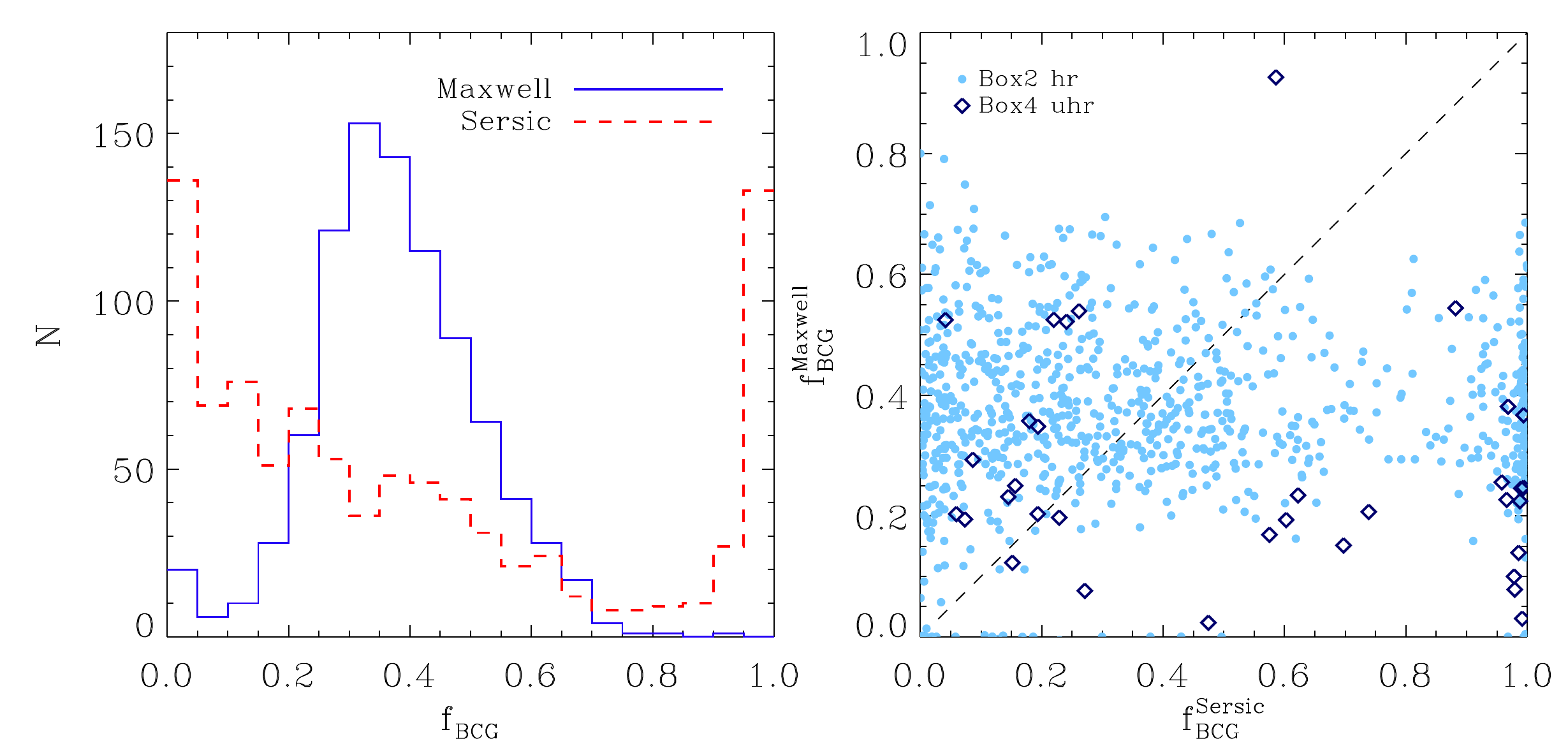}
  \caption{{\textbf{Left panel:}}
  Histogram of the mass fraction ascribed to the BCG according to the
  double-Maxwell fits (blue) and the double-S{\'e}rsic fits (red).
  The mass ascribed to the BCG is always the mass of the slower
  (Maxwellian fits) or the innermost (S{\'e}rsic fits) component.
  \textbf{{Right panel:}}
  BCG mass fractions obtained from the Maxwell fits versus those
  obtained with the S{\'e}rsic fits.
  Light blue symbols show the clusters from Box2b, while dark blue open
  diamonds mark the galaxy groups selected from Box4.
  There is no evident correlation between the mass partitioning obtained
  with the two different methods, and there is also no difference
  between galaxy groups and clusters.
  }
  {\label{fig:massfrac_bcg}}
\end{figure}

The left panel of Figure~\ref{fig:massfrac_bcg} shows a histogram of
the mass fractions of the BCG, $f_\mathrm{BCG}$, obtained with both
methods for all 928 halos.
The blue line shows the distribution found from the double-Maxwell
fits, while the red line shows the fractions obtained from the
double-S{\'e}rsic fits.
The right panel of the same figure shows a scatter plot of the BCG
mass fractions obtained with both methods.
As can clearly be seen, there is no correlation at all between the
mass fractions resulting from the two methods:
while~there is only a small amount of clusters that have BCG mass
fractions below 10\% and none with BCG mass fractions above 90\%
according to the double-Maxwell method, the double-S{\'e}rsic method
results in about half of the clusters having BCG mass fractions
of about 0 or 100\%, clearly indicating that in those cases a
double-S{\'e}rsic fit is not necessary and their radial surface
density profiles can be well described by a single S{\'e}rsic profile.

The latter is in good agreement with observations of radial surface
brightness profiles for massive elliptical galaxies, where both single-
and double-S{\'e}rsic-profiles are observed without a~clear correlation
to the global dynamical state of the cluster.
From the double-Maxwell method, we find BCG mass fractions generally
ranging between $20\%<f_\mathrm{BCG}<70\%$, but the large mass
fractions are rare and most of the BCGs have mass fractions
between 30 and 40\%, which is in agreement with observational
fractions obtained for the BCGs in very massive clusters (e.g.,
Presotto et al., 2014~\cite{presotto:2014},
\mbox{Burke et al., 2015~\cite{burke:2015}}).

\textls[-5]{We use the BCG mass fractions obtained with both methods
to decide whether a double-component fit is needed for the velocity
distributions and the surface density profiles or if a single-component
fit is sufficient:}
if the BCG mass fraction obtained through a double-Maxwell fit is
below $f_\mathrm{BCG}=10\%$ or~above $f_\mathrm{BCG}=90\%$, we judge
that there is no clear signal of a second component in this fit and
we thus classify these clusters as single-Maxwell clusters.
If the BCG mass fraction is between these values, i.e.,
$10\%<f_\mathrm{BCG}<90\%$, we gauge the double-Maxwell-fit to be
necessary and thus classify the cluster as double-Maxwell cluster.
As shown in the upper part of Table~\ref{tab:fractions}, the fraction
of single-Maxwell clusters is below $6\%$, and nearly all clusters
show velocity distributions that reflect two-component systems.
Therefore, we conclude that the typical galaxy cluster shows a
two-component behaviour in its velocity distribution, in agreement
with recent observations---for~example, by
\mbox{Longobardi et al., 2015~\cite{longobardi:2015}} and
Bender et al., 2015~\cite{bender:2015}.

\begin{table}[t]
  \caption{Relative fractions of the 928 clusters and groups with regard
  to their Maxwell- and S{\'e}rsic-fit~properties.}
  \centering
  \begin{tabular}{lcc}
    \toprule
    &
    {\boldmath $N_\mathrm{cluster}$} & 
    {\boldmath $f_\mathrm{cluster}~(\%)$} \\
    \midrule
    Single Maxwell sufficient    & ~~53 & ~~5.7 \\
    Double Maxwell needed        &  875 &  94.3 \\
    Single S{\'e}rsic sufficient &  386 &  41.6 \\
    Double S{\'e}rsic needed     &  542 &  58.4 \\
    \midrule
    Single Maxwell, Single S{\'e}rsic (s/s) & ~~29 & ~~3.1 \\
    Double Maxwell, Single S{\'e}rsic (d/s) &  357 &  38.5 \\
    Single Maxwell, Double S{\'e}rsic (s/d) & ~~24 & ~~2.6 \\
    Double Maxwell, Double S{\'e}rsic (d/d) &  518 &  55.8 \\
    \bottomrule
  \end{tabular}
  \label{tab:fractions}
\end{table}

Similarly, we classify a galaxy cluster as a single-S{\'e}rsic cluster
if the BCG mass fraction obtained from the double-S{\'e}rsic fit is
below $f_\mathrm{BCG}=10\%$ or above $f_\mathrm{BCG}=90\%$, while
we classify a cluster as double-S{\'e}rsic cluster if the BCG mass
fraction is between $10\%<f_\mathrm{BCG}<90\%$.
Here, we clearly see the same split-up that we already saw from
Figure~\ref{fig:massfrac_bcg}, i.e., that about half of the
clusters are single-S{\'e}rsic clusters while the other half are
double-S{\'e}rsic clusters, with a slight trend towards the latter
\mbox{(see upper part of Table~\ref{tab:fractions})}.

\begin{figure}[t]
  \centering
  \includegraphics[width=0.5\textwidth]{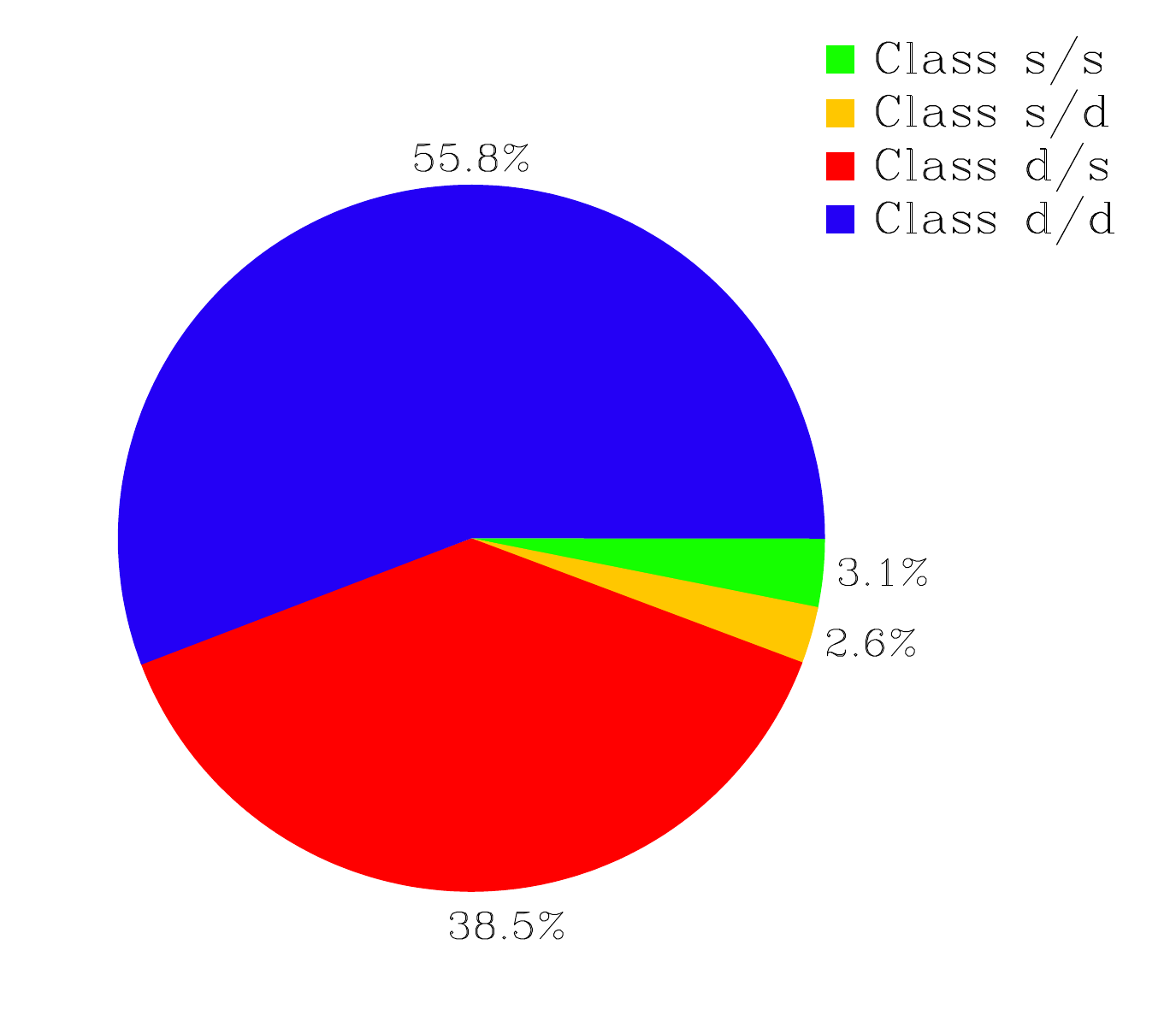}
  \caption{Fraction of galaxy clusters and groups that can be
  described best by
  a single Maxwell distribution and a single S{\'e}rsic profile
  (class s/s; green),
  a double Maxwell distribution and~a~single S{\'e}rsic profile
  (class d/s; red),
  a single Maxwell distribution and~a~double~S{\'e}rsic profile
  \mbox{(class s/d; yellow)},
  and a double Maxwell distribution and a double S{\'e}rsic profile
  (class d/d; blue).
  }
  {\label{fig:piechart}}
\end{figure}

Using both classifications, we can now test how many clusters show
a double-fit-behaviour in~both the velocity distribution and the
surface density profiles.
We find that this is the case for more than half of the galaxy clusters
in our sample, as shown in the lower part of Table~\ref{tab:fractions}
and Figure~\ref{fig:piechart}, while~about 40\% of the clusters are
double-Maxwell but single-S{\'e}rsic clusters.
The single-Maxwell clusters represent less than 6\% of all our
clusters; they are roughly evenly distributed between single-S{\'e}rsic
and double-S{\'e}rsic~cases.

From these results, we conclude that the velocity distribution of
a cluster can still distinguish between the component that belongs
to the direct potential of the BCG and the outer component that was
accreted onto the cluster and stored in the outer regions of the BCG
through stripping and flyby events, building up the DSC component
that still retains this memory of the assembly history.
On the other hand, the imprint of this assembly history is not always
visible in the radial surface density profiles of the cluster BCGs
as a separate component, where only in some cases the BCG can be
separated from the DSC through the surface density profiles, while,
in other cases, this is not possible.
Whether the assembly history can be traced from the shape of the
outer stellar halo radial density profiles of BCGs and galaxies
in general will be part of a forthcoming study (see Remus et al.,
2016~\cite{remus:2016} for a preview on these results).

\section{Mass--Velocity-Dispersion Relation}

\begin{figure}[t]
\centering
  \includegraphics[width=1\textwidth]{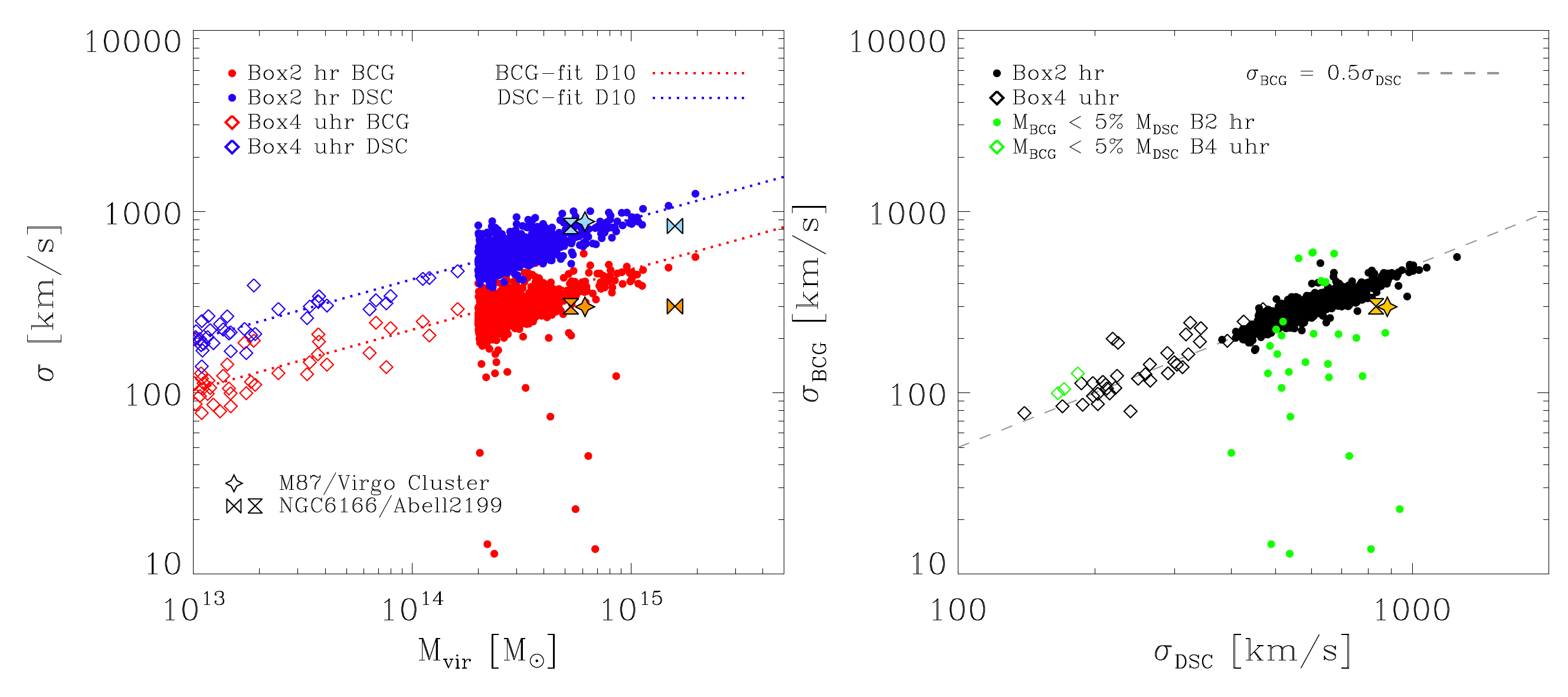}
  \caption{\textbf{{Left panel:}}
  Velocity dispersion $\sigma$ obtained from the double-Maxwellian
  fit versus virial mass $M_\mathrm{vir}$
  for the BCG-component (red)
  and the DSC-component (blue)
  for all galaxy clusters in Box2b (filled~circles)
  and all galaxy groups in Box4 (open diamonds).
  The red and blue lines are not fitted to the data presented here but
  are those from Dolag et al., 2010~\cite{dolag:2010}.
  The orange and light blue symbols show velocity dispersions from
  observations of {NGC}~6166 in the cluster Abell~2199 (bowties/hourglasses)
  from Bender et al., 2015~\cite{bender:2015}
  and of M~87 in the Virgo cluster (stars)
  from Longobardi et al., 2015~\cite{longobardi:2015}.
  The~virial mass for Virgo is taken from Tully, 2015~\cite{tully:2015}
  using the average of the masses based on the total K-band luminosity
  and the virial mass inferred from the zero velocity surface.
  For Virgo,~this~is in agreement with {measurements by}
  {\footnotesize PLANCK}~\cite{planckmission:2016};
  for Abell~2199, we included measurements of the virial mass from the
  {\footnotesize PLANCK} mission~\cite{planckmission:2016} (hourglasses)
  as they give significantly smaller values than the one inferred
  by Tully, 2015~\cite{tully:2015} (bowties).
  \textbf{Right panel:}
  BCG~velocity dispersion $\sigma_\mathrm{BCG}$ versus DSC~velocity
  dispersion $\sigma_\mathrm{DSC}$ obtained from the double-Maxwellian
  fits for the same clusters (filled circles) and groups (open diamonds)
  as in the left panel.
  The grey dashed line shows the $\sigma_\mathrm{BCG} =
  0.5~\sigma_\mathrm{DSC}$ relation.
  Green symbols mark all clusters and groups for which the BCG-component
  has less than~$5\%$ of the stellar mass of the total stellar mass
  of the system, according to the mass partitioning obtained from the
  double-Maxwellian fit.
  The yellow symbols show the same observations as in the left panel.
  }
  {\label{fig:m_sig}}
\end{figure}

Finally, we want to see if a correlation exists between the velocity
dispersion obtained from~the~Maxwell fits for the BCG and the DSC and
the virial mass of the host cluster, as presented
by~\mbox{Dolag et al., 2010~\citep{dolag:2010}}.
For this purpose, the left panel of Figure~\ref{fig:m_sig} shows the
velocity dispersion of~the~BCG component versus the virial mass of
the cluster in red, and the velocity dispersion of~the~DSC versus
the virial mass of the cluster in blue.
As can clearly be seen, we find a strong correlation for~both
components with the virial mass of the cluster, and these correlations
hold even for the galaxy groups in the lower mass regime, indicating
that the split-up between the brightest group galaxies (BGGs) and the
Intra-group light (IGL) behaves similarly to that of clusters, clearly
hinting at a similar growth mechanism for the IGL through stripping.

The left panel of Figure~\ref{fig:m_sig} also shows the fits to the
velocity-dispersion--virial-mass relation presented by Dolag et al.,
2010~\citep{dolag:2010} as red and blue dotted lines for the BCGs
and the DSC, respectively.
Although not fitted to the current simulation set but obtained from
a less advanced simulation set of the local universe, these relations
perfectly describe the behaviour found for the Magneticum simulation
sample of galaxy clusters and galaxy groups, even at the low mass end.
This additionally proves that this behaviour is independent of
the details of the subgrid models included in the simulations.

As can also be seen from Figure~\ref{fig:m_sig}, the relation between
the velocity dispersion and the virial mass for the BCGs and the DSC
has the same slope, with the DSC simply having overall larger velocity
dispersions than the BCGs.
More precisely, as shown in the right panel of Figure~\ref{fig:m_sig},
the relation between the velocity dispersions of the BCG and the DSC
is very tight and can be described as
\begin{equation}
\sigma_\mathrm{BCG} = 0.5~\sigma_\mathrm{DSC}.
\end{equation}
Again, this behaviour holds even at the galaxy group mass scale,
as indicated by the open diamonds marking the groups selected from
the smaller volume Box4 with the higher resolution.
In~addition, this not only demonstrates that galaxy groups and clusters
show a similar behaviour, but it also proves that the correlations
presented here are independent of
the resolution of the simulation and thus only driven by physical
processes like accretion and star formation.

Interestingly, we can also explain the few outliers that can be
seen in both the velocity-dispersion--\allowbreak virial-mass relation
and the velocity-dispersion relation between the BCGs and their DSC:
if we mark all clusters (and groups) where the BCG mass fraction
obtained from the double-Maxwellian fit is below~5\%
(green circles and diamonds in the right panel of Figure~\ref{fig:m_sig}),
all~outliers are captured.
This clearly indicates that for all galaxy clusters where a
double-Maxwellian fit is the better representation of the velocity
distribution, the discussed relation between both components and the
virial mass of the cluster is present and very tight, and driven by
the assembly history of~the~clusters.

In addition, we also included the observations
for {NGC}~6166 in the cluster Abell~2199 from
\mbox{Bender et al., 2015~\cite{bender:2015}}
and M~87 in the Virgo cluster from
Longobardi et al., 2015~\cite{longobardi:2015}
in both panels of Figure~\ref{fig:m_sig}
(for details on the virial mass estimated for these clusters,
see the figure caption).
Both~observations are in excellent agreement with the correlations
found in this study, especially with respect to the BCG--DSC
velocity dispersion correlation.

\section{Discussion and Conclusion}

In this work, we presented a detailed and statistically sound analysis
of the stellar velocity distributions and the projected stellar radial
surface density profiles of galaxy clusters and galaxy groups selected
from the Magneticum pathfinder simulation sample.
Using two volumes of different sizes and resolutions, we showed that
for more than $90\%$ of all 928 clusters and groups in our sample the
velocity distributions are represented best by a superposition of two
Maxwellian distributions, with the slower component representing the
BCG of the cluster and the faster component representing the DSC.
We demonstrated that the relative mass fractions of the BCGs found
through these fits is~in~agreement with recent observations.
This behaviour in the velocity distribution strongly supports the idea
that the DSC is built up from stripping of smaller satellites within
the cluster potential close to~its center, where the BCG resides.

Furthermore, we found that there is a clear and tight correlation
between the velocity dispersions of the two components obtained by
these fits and the virial mass of the host clusters, and that this
correlation holds down to group-mass scales.
We also demonstrated that the velocity dispersions of~both components
are correlated tightly, with the BCG having about half the velocity
dispersion of~the DSC, and that the few available observations that
distinguish between both components are~in~excellent agreement with
our results.

Additionally, we tested if the same separation into two distinct
components is reflected in the projected radial surface density
profiles of~the~cluster.
Interestingly, we could only find a~separation of~the radial profiles
into two components in about half of the clusters that exhibit a
double-Maxwell imprint in the velocity distribution, clearly showing
that the radial profile is not always suitable for distinguishing the
two components and that further indicators are needed in the radial
stellar profiles to obtain information about the assembly history of
galaxy clusters.
This issue will be addressed in~a~forthcoming study.

Lastly, we also tested if we can find a correlation between the
velocity distribution behaviour of~the~DSC and the X-ray or shock
properties in the four objects we have examined in detail.
No such correlations were evident.
On the contrary, we find indications that the X-ray and shock
properties describe the very recent assembly history of the cluster,
where the presence of shocks and X-ray offsets indicate an ongoing
merger event, while the velocity distribution of the stellar component
of the galaxy clusters and the BCGs appears to be an indicator for
the earlier assembly history of the cluster.

\vspace{6pt}
\acknowledgments{We thank the anonymous referees for their helpful
comments.
The Magneticum Pathfinder simulations were partly performed at the
Leibniz-Rechen\-zentrum with {CPU} time assigned to the Project ``pr86re''.
This work was supported by the {DFG} Cluster of Excellence ``Origin
and Structure of the Universe''.
We are especially grateful for the support by {M.}~Petkova through
the Computational Center for Particle and~Astrophysics~(C2PAP).}

\authorcontributions{K.D. performed the simulations;
R.-S.R. and T.L.H. analysed the data;
R.-S.R. wrote the~paper.}

\conflictsofinterest{The authors declare no conflict of interest.}


\end{document}